\documentclass[sn-mathphys,Numbered]{sn-jnl}

\usepackage{graphicx}%
\usepackage{multirow}%
\usepackage{amsmath,amssymb,amsfonts}%
\usepackage{amsthm}%
\usepackage{mathrsfs}%
\usepackage[title]{appendix}%
\usepackage{xcolor}%
\usepackage{textcomp}%
\usepackage{manyfoot}%
\usepackage{booktabs}%
\usepackage{algorithm}%
\usepackage{algorithmicx}%
\usepackage{algpseudocode}%
\usepackage{listings}%
\usepackage{array}
\usepackage{colortbl}

\theoremstyle{thmstyleone}%
\theoremstyle{thmstyletwo}%

\theoremstyle{thmstylethree}%

\begin{document}

\title[Efficient Information Reconciliation for High-Dimensional Quantum Key Distribution]{Efficient Information Reconciliation for High-Dimensional Quantum Key Distribution}


\author*[1]{\fnm{Ronny} \sur{Mueller}}\email{ronmu@dtu.dk}

\author[2,3]{\fnm{Domenico} \sur{Ribezzo}}

\author[1]{\fnm{Mujtaba} \sur{Zahidy}}

\author[1]{\fnm{Leif Katsuo} \sur{Oxenl{\o}we}}

\author[4]{\fnm{Davide} \sur{Bacco}}

\author[1]{\fnm{S{\o}ren \sur{Forchhammer}}}

\affil*[1]{\orgdiv{Department of Electrical and Photonics Engineering}, \orgname{Technical University of Denmark},  \orgaddress{\city{Lyngby}, \country{Denmark}}}

\affil[2]{\orgname{National Institute of Optics of National Research Council,}, \orgaddress{\city{Florence}, \country{Italy}}}

\affil[3]{\orgname{University of Naples Frederico II}, \orgaddress{\city{Naples}, \country{Italy}}}

\affil[4]{\orgdiv{Department of Physics and Astronomy}, \orgname{University of Florence}, \orgaddress{\city{Florence}, \country{Italy}}}


\abstract{The Information Reconciliation phase in quantum key distribution has significant impact on the range and throughput of any QKD system. We explore this stage for high-dimensional QKD implementations and introduce two novel methods for reconciliation. The methods are based on nonbinary LDPC codes and the Cascade algorithm, respectively,  and achieve efficiencies close the the Slepian-Wolf bound on q-ary symmetric channels.}




\maketitle

\section{Introduction}\label{sec1}

Quantum Key Distribution (QKD) protocols allow for secure transmission of information between two entities, Alice and Bob, by distributing a symmetric secret key via a quantum channel \cite{BEN84, Pirandola_2020}. The process involves a quantum stage where quantum information is distributed and measured. This quantum stage is succeeded by post-processing. In this purely classical stage, the results of the measurements undergo a reconciliation process to rectify any discrepancies before a secret key is extracted during the privacy amplification phase. The emphasis of this work is on the phase of information reconciliation, which  impacts the range and throughput of any QKD system.

Despite the considerate development of QKD technology using binary signal forms, its high-dimensional counterpart (HD-QKD)\cite{201900038}  has seen significantly less research effort so far. However, HD-QKD offers several benefits, including higher information efficiency and increased noise resilience \cite{PhysRevA.61.062308, PhysRevA.94.052323, PhysRevX.9.041042, Cozzolino_2019}. Although the reconciliation phase for binary-based QKD has been extensively researched, little work has been done to analyse and optimize this stage for HD-QKD, apart from introducing the layered scheme in 2013 \cite{6502993}. This study addresses this research gap by introducing two novel methods for information reconciliation for high-dimensional QKD and analysing their performance.

Unlike the majority of channel coding applications, the (HD)-QKD scenario places lesser demands on latency and throughput while emphasizing significantly the minimization of information leakage. Spurred by this unique setting, the strong decoding performance of nonbinary LDPC codes \cite{706440}, and their inherent compatibility with high dimensions, we investigate the construction and utilization of nonbinary LDPC codes for post-processing in HD-QKD protocols as the first method.

The second method we investigate is the Cascade protocol \cite{10.1007/3-540-48285-7_35}. It is one of the earliest proposed methods for reconciling keys. While many rounds of communication required by Cascade and concerns about resulting limitations on throughput have led to a focus on syndrome-based methods \cite{blind, Kiktenko_2021, Martinez-Mateo2013} in the past decade, recent research has shown that sophisticated software implementations can enable Cascade to achieve high throughput even with realistic latency on the classical channel \cite{mao2021high, pedersen2013high}. Motivated by these findings, we explore the usage of Cascade in the reconciliation stage of HD-QKD and propose a modification that enables high reconciliation efficiency for the respective quantum channel.

To the best of our knowledge, the only prior work investigating information reconciliation for high-dimensional QKD is the aforementioned layered scheme, introduced in 2013. The layered scheme is based on decoding bit layers separately using $\lceil \log_2(q)\rceil$ binary LDPC codes. It is similar in concept to the multilevel coding and multistage decoding methods used in slice reconciliation for continuous-variable (CV) QKD \cite{1266817}.

LDPC codes have been widely studied and optimized for the use in binary QKD systems \cite{Mao2019, 5205475, PhysRevApplied.8.044017, reviewer_bin_ir} and can reach good efficiency with high throughput. The use of nonbinary LDPC codes has also been investigated for binary QKD \cite{5649550}. Modifications of Cascade have been shown to achieve efficiency performance close to the theoretical limit \cite{pacher} on binary QKD-systems while simultaneously reaching high throughput \cite{mao2021high}. Except for the layered scheme, neither LDPC codes, Cascade, or any other error correction method has yet been optimized, modified or analysed for the use in high-dimensional QKD.

This work has the following outline: In Sec. \ref{sec:ir}, the general scenario of Information Reconciliation is introduced. This is followed by an introduction of Nonbinary LDPC codes and Density Evolution in Sec. \ref{sec:nb_codes_background}. In Sec. \ref{sec:cascade_intro},  the usage of Cascade is reviewed and a novel algorithm for high-dimensional Information Reconciliation, high-dimensional Cascade, is introduced. The results, i.e. the performance of both Nonbinary LDPC codes and high-dimensional Cascade, are shown  in Sec. \ref{sec:results} followed by a discussion and comparison in Sec. \ref{sec:dis}. The work concludes in a short review of the achieved results in Sec. \ref{sec:con}.

\section{Background}\label{sec2}

In this Section, we describe the general setting and channel model, and introduce relevant figures of merit. We then continue to describe the two proposed methods, nonbinary LDPC codes and high-dimensional Cascade,  in more detail.

\subsection{Information reconciliation}\label{sec:ir}

The goal of the information reconciliation stage in QKD is to correct any discrepancies between the keys of the two parties while minimizing the information leaked to potential eavesdroppers. Generally, Alice sends a random string $\mathbf{x}=(x_0,...,x_{n-1})$, $x_i = {0,...,q-1}$ of $n$ qudits of dimension $q$ to Bob, who measures them and obtains his version of the string $\mathbf{y}=(y_0,...,y_{n-1})$, $y_i = {0,...,q-1}$. We assume that the quantum channel can be accurately represented by a substitute channel where $\mathbf{x}$ and $\mathbf{y}$ are correlated as a $q$-ary symmetric channel since errors are typically uncorrelated and symmetric. The transition probabilities of such a channel are as follows:

\begin{equation}
\text{P}(y_i|x_i) = \begin{cases} 1-p & y_i=x_i,\\
\frac{p}{q-1} & \text{else}.
\end{cases}
\end{equation}

\vspace{2mm}
\noindent Here, the parameter $p$ represents the channel transition probability. We refer to the symbol error rate between $\mathbf{x}$ and $\mathbf{y}$ as the quantum bit error rate (QBER) in a slight abuse of notation but consistent with experimental works on HD-QKD. In our simulations, we assume the QBER to be an inherent channel property, making it equivalent to the channel parameter $p$. In addition to the qudits, Alice also sends classical messages, e.g. syndromes or parity bits, which are assumed to be error-free. From a coding perspective, this is equal to asymmetric Slepian-Wolf coding with side information at the receiver, where the syndrome $\mathbf{s}$ represents the compressed version of $\mathbf{x}$, and $\mathbf{y}$ is the side information. A more detailed explanation of this equivalence can be found in \cite{1042242}. For an interpretation of Cascade in the context of linear block codes see \cite{pacher}. Any information leaked to a potential eavesdropper at any point during the quantum key distribution must be subtracted from the final secret key during privacy amplification \cite{pa_org}. The information leaked during the information reconciliation stage will be denoted by leak${}_{\text{IR}}$. In the case of LDPC codes, assuming no rate adaptation, it can be upper-bounded by the syndrome length in bits, $\text{leak}_{\text{IR}} \leq m$, with $m$ being the length of a binary representation of the syndrome string. In the case of Cascade, it can be upper-bounded by the number of parity bits sent from Alice to Bob \cite{Lo_2003}, although attention has to be paid to special cases in relation to the parameter estimation phase of QKD post-processing \cite{tupkary2023using}. Using the Slepian-Wolf bound \cite{1055037}, the minimum amount of leaked information required to successfully reconcile with an arbitrarily low failure probability in the asymptotic limit of infinite length is given by the conditional entropy:

\begin{equation} \label{upper_bound_ir}
\text{leak}_{\text{IR}} \geq n\text{H}(X|Y).
\end{equation}

\noindent The conditional entropy (base $q$) of the $q$-ary symmetric channel, assuming independent and identically distributed input $X$, can be expressed as
\begin{equation}
\text{H}(X|Y) = -((1-p)\text{log}_q(1-p) + p\cdot \text{log}_q(\frac{p}{q-1})).
\end{equation}

\noindent A code's performance in terms of relative information leakage can be measured by its efficiency $f$, given by

\begin{equation}
f = \frac{\text{leak}_{\text{IR}}}{n\text{H}(X|Y)}.
\end{equation}

\noindent It is important to note that an efficiency of $f>1$ corresponds to leaking more bits than required by the theoretical minimum of $f=1$, which represents the best possible performance according to the Slepian-Wolf bound. In practice, systems have $f>1$ due to the difficulty of designing optimal codes, finite-size effects, and the inherent trade-off between efficiency and throughput. For more details on achievable information leakage, including respect to finite-size effects, see for example \cite{Tomamichel_2017}. In the following sections, we restrict ourselves to $q$ being a power of 2. Both approaches can function without this restriction, but it allows for more efficient implementation of the reconciliation and is commonly seen in physical implementations of the quantum stage due to symmetries.

The Information Reconciliation phase is succeeded by an Error Verification stage wherein an estimate of the expected probability of correctness is obtained. Here, correctness refers to the agreement of both Alice's and Bob's versions of the key after information reconciliation. In practical terms, this is frequently accomplished through the exchange and comparison of hashes \cite{Luetkenhaus_1999, Fung_2010}. In case of a disagreement, the error correction can be repeated with the cost of doubling the leaked information. If this is not feasible, e.g. the additional leakage prohibits any secret key extraction, the keys of this round are discarded.

\begin{figure}
    \centering
    \includegraphics[width=0.5\linewidth]{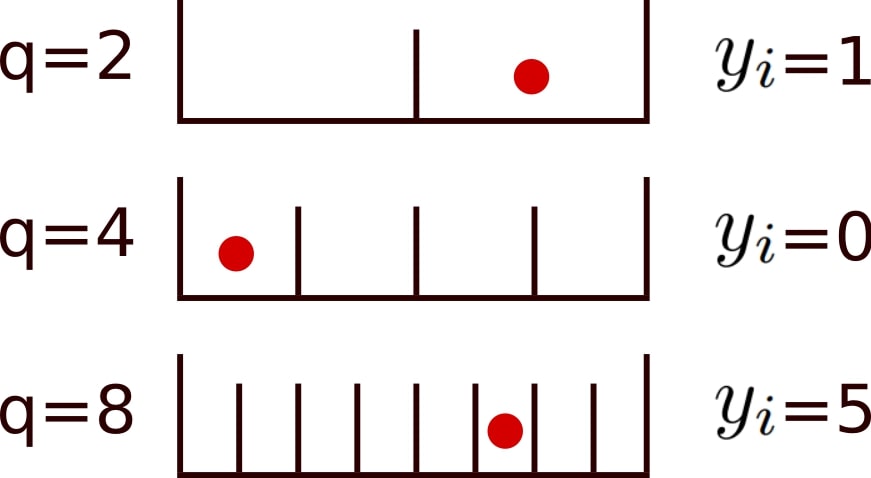}
    \caption{Example of a qudit using a time-bin implementation \cite{Boaron_2018}. The dimension of the qudit is set by the number of bins grouped together, while the value is determined by the measured arrival time.}
    \label{fig:scheme}
\end{figure}

\subsection{Nonbinary LDPC codes}
\subsubsection{Codes \& Decoding}

We provide here a short overview over nonbinary LDPC codes and their decoding based on the concepts and formalism of binary LDPC codes. For a comprehensive review of those, we refer to \cite{ldpcbook}.

Nonbinary LDPC codes can be described by their parity check matrix $\mathbf{H}$, with $m$ rows and $n$ columns, containing elements in a Galois Field (GF) of order $q$. To enhance clarity in this section, all variables representing a Galois field element will be marked with a hat, for instance, $\hat{a}$. Moreover, let $\oplus, \ominus, \otimes,$ and $\oslash$ denote the standard operations on Galois field elements: Addition, subtraction, multiplication, and division \cite{multis}.
An LDPC code can be depicted as a bipartite graph, known as the Tanner graph. In this graph, the parity-check equations form one side, called check nodes, while the codeword symbols represent the other side, known as variable nodes. The Tanner graph of a nonbinary LDPC code also has weighted edges between check and variable nodes, where each weight corresponds to the respective entry of $\mathbf{H}$. The syndrome $\mathbf{s}$ of the $q$-ary string $\mathbf{x}$ is computed as $\mathbf{s} = \mathbf{H}\mathbf{x}$.

For decoding, we employ a log-domain FFT-SPA \cite{6577118, 1312606}. In-depth explanations of this algorithm can be found in \cite{4787626, 6987278}, but we provide a summary here for the sake of completeness. Let $Z$ represent a random variable taking values in GF($q$), such that $\text{P}(Z_i = k)$ indicates the probability that qudit $i$ has the value $k=0,...,q-1$. The probability vector $\mathbf{p}=(p_0,...p_{q-1})$, $p_j = \text{P}(Z=j)$, can be converted into the log-domain using the generalized equivalent of the log-likelihood-ratio (LLR) in the binary case, $\mathbf{m}=(m_0,...,m_{q-1})$, $m_j = \text{log}\frac{\text{P}(Z=0)}{\text{P}(Z=j)} = \log(\frac{p_0}{p_j})$. Unless specified otherwise, the logarithm is taken to base $e$. Given the LLR representation, probabilities can be retrieved through $p_j = \exp(-m_j)/\sum_{k=0}^{q-1} \exp(-m_k)$. We use $p(\cdot)$ and $m(\cdot)$ to denote these transforms. To further streamline notation, we define the multiplication and division of an element $\hat{a}$ in GF$(q)$ and an LLR message as a permutation of the indices of the vector:

\begin{align}
\hat{a} \cdot \mathbf{m} &:= (m_{\hat{0} \otimes \hat{a}},...,m_{\hat{q-1} \otimes \hat{a}})\\
\mathbf{m} / \hat{a} &:= (m_{\hat{0}\oslash \hat{a}},...,m_{\hat{q-1} \oslash \hat{a}}),
\end{align}

\noindent where the multiplication and division of the indices occur in the Galois Field. These permutations are necessary as we need to weigh messages according to their edge weight during decoding. We further define two transformations involved in the decoding,

\begin{align}
\Bar{\mathcal{F}}(\mathbf{m}, \hat{H}{ij}) &= \mathcal{F}(p(\hat{H}{ij}\cdot \mathbf{m}))\\
\Bar{\mathcal{F}}(\mathbf{m}, \hat{H}{ij})^{-1} &= m(\mathcal{F}^{-1}(\mathbf{m}))/\hat{H}{ij},
\end{align}

\noindent where $\mathcal{F}$ represents the discrete Fourier transform. Note that for $q$ being a power of 2, the Fast Walsh Hadamard Transform can be utilized. The decoding process then consists of two iterative message-passing phases, from check nodes to variable nodes and vice versa. The message update rule at iteration $l$ for the check node to variable node message corresponding to the parity check matrix entry at $(i,j)$ can be expressed as

\begin{equation}
    \mathbf{m}^{(l)}_{ij,\text{CV}} = \mathcal{A}(\hat{s}^{\prime}_i) \Bar{\mathcal{F}}^{-1}( \underset{j\prime \in \mathcal{M}(i)/j}{\Pi} \Bar{\mathcal{F}}(\mathbf{m}^{(l-1)}_{ij\prime}, \hat{H}_{ij\prime}), \hat{H}_{ij}),
\end{equation}

\noindent where $\mathcal{M}(i)$ denotes the set of all check nodes in row $i$ of $\mathbf{H}$. The matrix $\mathcal{A}$, defined as $\mathcal{A}_{kj}(\hat{a}) = \delta( \hat{a} \oplus k \ominus j) - \delta(a\ominus j)$, accounts for the possible occurrence of nonzero syndromes \cite{6987278}. The weighted syndrome value is calculated as $\hat{s}^{\prime}_i = \hat{s}_i \oslash \hat{H}_{ij}$. The a posteriori message of column $j$ can be written as
\begin{equation}
    \Tilde{\mathbf{m}}^{(l)}_j = \mathbf{m}^{(0)}(j) + \underset{i^{\prime} \in \mathcal{N}(j)}{\sum} \mathbf{m}^{(l)}_{i^{\prime}j,\text{CV}},
\end{equation}
where $\mathcal{N}(j)$ is the set of all check nodes in column $j$ of $\mathbf{H}$. The best guess $\Tilde{\mathbf{x}}$ at each iteration $l$ can be calculated as the minimum value of the a posteriori, $\Tilde{x}_j^{(l)} = \text{argmin} (\Tilde{\mathbf{m}}^{l}_j)$. The second message passings, from variable to check nodes, are given by
\begin{equation}
    \mathbf{m}_{ij, \text{VC}}^{(l)} = \Tilde{\mathbf{m}}^{(l)}_j - \mathbf{m}^{(l)}_{ij, \text{CV}}.
\end{equation}

\noindent The message passing continues until either $\mathbf{H}\Tilde{\mathbf{x}} = \mathbf{s}$ or the maximum number of iterations is reached.

To allow for efficient reconciliation for  different QBER values, a rate-adaptive scheme is required. We use the blind reconciliation protocol \cite{blind} as it has a better performance with respect to efficiency than direct rate-adaption \cite{Elkouss_2010_rate_adaptive}. A fixed fraction $\delta$ of symbols is chosen to be punctured or shortened. Puncturing refers to replacing a key bit with a random bit that is unknown to Bob. For shortening, the value of the bit is additionally sent to Bob over the public channel. Puncturing, therefore, increases the code rate, while shortening lowers it. The rate of a code with $p$ punctured and $s$ shortened bits is then given by

\begin{equation}
    R = \frac{n-m-s}{n-p-s}.
\end{equation}

\noindent To see how rate adaption influences the bounding of $\text{leak}_{\text{IR}}$, see \cite{5650099}. The blind scheme introduces interactivity into the LDPC reconciliation. Given a specific code, we start out with all bits being punctured and send the respective syndrome to Bob. Bob attempts to decode using the syndrome. If decoding fails, Alice transforms  $\lceil n(0.028 - 0.02R)\rceil$ \cite{symm_blind} punctured bits into shortened bits, and resends the syndrome. This value is a heuristic expression and presents a trade-off between the number of communication rounds and the efficiency. Bob tries to decode again and requests more bits to be shortened in case of failure. If there are no punctured bits left to be turned into shortened bits, Alice reveals plain key bits instead. This continues until either decoding succeeds or  the whole key is revealed by Alice successively sending all key bits through the public channel.

\subsubsection{Density Evolution}
In the case of a uniform edge weight distribution, the asymptotic decoding performance of LDPC codes for infinite code length is entirely determined by two polynomials \cite{4712621, 910577}:

\begin{equation} \label{lamda}
    \lambda(x) = \sum_{i=0}^{d_{\text{v, max}}} \lambda_i x^{i-1} \quad \rho(x) = \sum_{i=0}^{d_{\text{c, max}}} \rho_i x^{i-1}.
\end{equation}

\noindent In these expressions, $\lambda_i$ ($\rho_i$) represents the proportion of edges connected to variable (check) nodes with degree $i$, while $d_{\text{v,max}}$ ($d_{\text{c,max}}$) indicates the highest degree of the variable (check) nodes.
Given these polynomials, we can then define the code ensemble $\mathcal{E}(\lambda, \rho)$, which represents all codes of infinite length with degree distributions specified by $\lambda$ and $\rho$. The threshold $p_t(\lambda, \rho)$ of the code ensemble $\mathcal{E}(\lambda, \rho)$ is defined as the worst channel parameter (QBER) at which decoding remains possible with an arbitrarily small failure probability. This threshold can be estimated using Monte-Carlo Density Evolution (MC-DE), which is thoroughly described in \cite{inproceedings}. This technique repeatedly samples node degrees according to $\lambda$ and $\rho$, and draws random connections between nodes for each iteration. With a sufficiently large sample size, this simulates the performance of a cycle-free code. Note that MC-DE is particularly well suited for nonbinary LDPC codes, as the distinct edge weights aid in decorrelating messages \cite{inproceedings}. During the simulation, we track the average entropy of all messages \cite{inproceedings}. When it falls below a certain value, decoding is considered successful. If this does not occur after a maximum number of iterations, the evaluated channel parameter is above the threshold of $\mathcal{E}(\lambda,\rho)$. Utilizing a concentrated check node distribution (which is favorable according to \cite{910580}) and a fixed code rate, we can further simplify to $\mathcal{E}(\lambda)$. The threshold can then be employed as an objective function to optimize the code design, which is commonly achieved using the Differential Evolution algorithm \cite{de}.

\subsection{Cascade}\label{sec:cascade_intro}
\subsubsection{Binary Cascade} 

Cascade \cite{10.1007/3-540-48285-7_35} is one of the earliest schemes proposed for information reconciliation and has seen widespread use due to its simplicity and high efficiency. Successive works on the original Cascade protocol have been trying to increase its performance by either substituting the parity exchange with error-correction methods \cite{winnow}, or by optimizing parameters like the top-level block sizes \cite{martinezmateo2014demystifying}. The binary Cascade protocol acting on a single frame can be summarised in the following steps:

\begin{itemize}
    \item Iteration 1:
    \begin{enumerate}
    \item The binary frame is divided into non-overlapping blocks of size $k_1$, where the value of $k_1$ usually depends on the estimated QBER of the frame. 
    \item Alice and Bob calculate the parity of each top-level-block and share them over a classical channel. 
    \item For those blocks where a mismatch between the parity of Alice and the parity of Bob is detected, a binary search is used to detect a single error. In general, iff a parity mismatches between Alice and Bob, a binary search can be performed on that block. The binary search consists of three steps:
    \begin{enumerate}
        \item Split the respective block in half.
        \item Calculate and exchange the parities of the two sub-blocks. One of the sub-blocks has a mismatching parity.
        \item  If the mismatching sub-block contains only 1 bit, the error is found. Otherwise, repeat Step (a) with the mismatching sub-block.
    \end{enumerate}
    \end{enumerate}
    \item Iteration $i$:
    \begin{enumerate}
    \item Apply a permutation on the frame and divide it into new top-level-blocks of size $k_i$. Repeat Steps 2 and 3 as described for the first iteration on the new blocks.
    \item Cascade step: Any erroneous bit detected in iteration $i$ also takes part in blocks created in previous iterations. After correcting those bits, their parity mismatches again and allows for the detection of another error using binary search. This error again takes part in blocks of all other iterations and can be used to detect more errors, creating the ''cascading" effect of the Cascade step.
    \end{enumerate}
\end{itemize}

\noindent The protocol stops after a fixed number of iterations. To the best of our knowledge, state-of-the-art in terms of efficiency with values up to $f=1.025$ is reached by a modification \cite{pacher} of the original Cascade. The additions of this modification constitute mainly of separating bits into groups of similar confidence and choosing optimal block sizes on those groups in Iteration 2. We will denote this version of Cascade as "binary Cascade" and it will be used for all comparisons. It has also been used as the base for a recent implementation reaching the highest throughout so far of $570$ Mbps \cite{mao2021high}.

\subsubsection{High-Dimensional Cascade}\label{hdc}
We propose the following modification to use Cascade for high-dimensional data, which we will denote by high-dimensional Cascade (HD-Cascade). We only highlight the differences compared to binary Cascade as described in \cite{pacher}. All the modifications we propose are additions to the binary algorithm and reduce back to binary Cascade for $q=2$ as they rely on correlations that only exist for $q>2$.

\begin{figure}
    \centering
    \includegraphics[width=\linewidth]{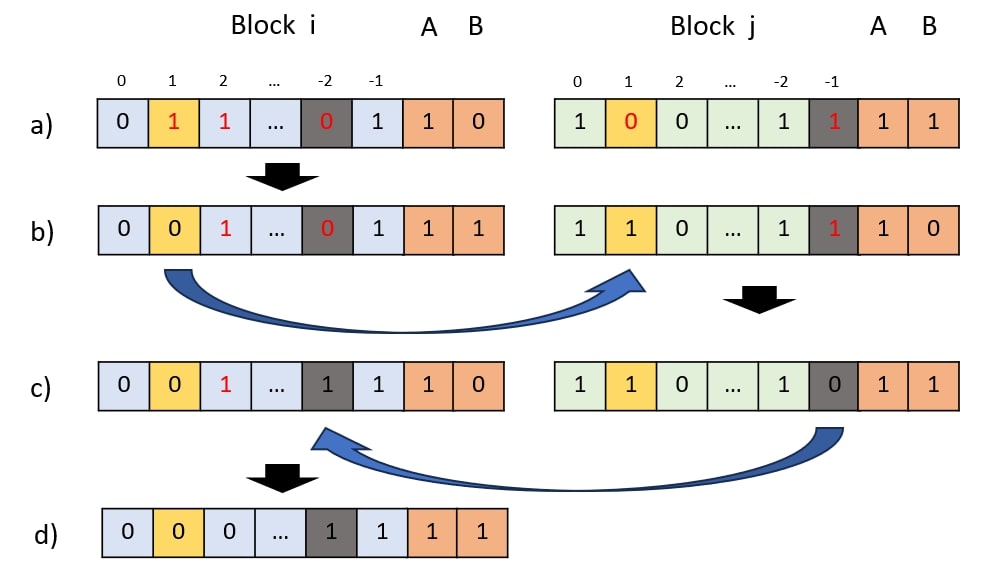}
    \caption{Example of the cascading step inside the first iteration of HD-Cascade. Blocks $i$ and $j$ are random top-level blocks in the first iteration. The bits at positions ($i,1$) \& ($j,1$) and ($i,-2$) \& ($j,-1$) originate from the same symbol, respectively. A and B denote the current parity of the block for Alice and Bob, respectively. a): Block $j$ has a matching parity, so no binary search is possible. Block $i$ has mismatching parity, a binary search reveals position 1 to be erroneous. b): After correcting the error, Bob's parity flips to match that of Alice. By requesting and correcting  the associated partner bit in block $j$, a mismatch in the parities is introduced. We can therefore run a binary search on block $j$ and detect the error at position ($j,-1$). c): By requesting and correcting the respective partner bit in block $i$, the parities mismatch again and allow for another round of binary search to reveal the error at position ($i,2$). d): All blocks have matching parities.}
    \label{fig:scheme}
\end{figure}

\begin{itemize}
    \item \textbf{Modification 1:} Initially, we map all symbols to an appropriate binary representation. Prior to the first iteration, we shuffle all bits while maintaining a record of which bits originate from the same symbol. We denote these bits as ''partner bits". This mapping effectively reduces the expected QBER used for block-size calculations in all iterations to $\text{QBER}_{\text{BIN}} = \frac{q}{2(q-1)} \text{QBER}_{\text{SYM}}$.\\
    
    \item \textbf{Modification 2:}     
    Upon detecting an error at any point during the protocol, immediately do the following:
    \begin{enumerate}
        \item Request all partner bits of the erroneous bit.
        \item If any of the partner bits are erroneous, the blocks they have been participating in now have mismatching parities again. Run a binary search on all mismatching blocks. Repeat Step 1 until no new errors are found.
    \end{enumerate}

     The conditional probability for a partner bit to be erroneous given the values of all previously transmitted bits for these bits is close to $1/2$. To be precise, it is equal to $1/2$ for bits that have not yet participated in any parity checks and then varies with the length of the smallest block they participated in \cite{pacher}. Note that this procedure allows for a cascading process in the first iteration already, and therefore requires all blocks to be processed sequentially followed by a Cascade step for each single error for maximum impact of the Cascade step.\\
    
    \item \textbf{Modification 3:} The fraction of errors corrected in the first iteration is significantly higher (often $>95\%$ in our simulations for high dimensions) compared to the binary version. This is due to the possibility of running a cascading process in the first iteration already, see Figure \ref{fig:scheme} for an example. Consequently, we need to increase the block sizes for the following iterations as the dimensionality increases, see Table~\ref{tab:table1}. The importance of partner bits increases with increasing dimension.
\end{itemize}

\subsubsection{Parallel High-Dimensional Cascade}\label{sec:phdc}

While the proposition in Section \ref{hdc} achieves great efficiency, it also requires all blocks to be processed serially. This results in a limited throughput, as a large amount of messages is required, i.e. a single message for every requested parity. We therefore propose the following adaptions for a more practical implementation of high-dimensional Cascade:

\begin{enumerate}
    \item \textbf{Modification 4:} In the first iteration, we split the binary representation into $v=\log_2(q)$ groups of size $n$; $n$ being the number of qudits per frame. Each group only contains bits of the same bit-plane, i.e. the first group contains all the first bits of all symbols, the second group contains all the second bits of all symbols and so on. We then apply permutations that are restricted to each group only.  After locating the error positions of one group using binary search, the corresponding partner bits will be located in all other groups. All blocks of one group can therefore be processed in parallel. When calculating the top-level block sizes of the following blocks, we adjust the expected QBER using the number of partner bits. It can be calculated for group $i$, $i=1,...,v$ as:
    \begin{equation}
        \text{QBER}_i = \text{QBER}_{\text{BIN}} - \frac{1}{2n}\sum_{j<i} \frac{\text{PB}_j}{v}, 
    \end{equation}

    where $\text{PB}_j$ denotes the number of partner bit requests originating from group $j$. To reduce the number of messages sent, we do not cascade on the partner bits until all groups have finished this first stage. The block creation for all other iterations follows the procedure described in \cite{pacher}.\\
    \item \textbf{Modification 5:} This modification replaces Modification 2 and the Cascade step of the binary protocol. After completing the binary search for all top-level blocks of an iteration, we collect all found errors into a list. We then do the following for the list of known errors:
    \begin{enumerate}
        \item All known error positions are fed into the Cascade step as a group and processed in parallel. For each error, locate the smallest block for which the error participates in a not-flagged iteration.
        \item Run a binary search on those smallest blocks that have a mismatch and locate the new error positions. Add the new errors to the list of all errors. Request the partner bits of these new errors if not already known and add them to the list of all known errors if they are erroneous. Flag the origin iterations of the blocks as already used for the respective input bits. The origin iteration of a block is that iteration in which the block has been created, either as a top-level block or as part of the binary search.
        \item Remove all errors from the list that have all iterations flagged. Correct all known errors and repeat Step (a). If there are no new errors to be found, the Cascade step terminates.
        
    \end{enumerate}
    
    \noindent The Cascade step can be stopped early in a  trade off between efficiency, throughput, and frame error rate.
    

\end{enumerate}

\section{Results}\label{sec:results}

\subsection{Nonbinary LDPC codes}\label{sec:nb_codes_background}

While the code-design and decoding techniques described above are feasible for any dimension $q$, we focus on $q = 4$ and $8$ as those are common in current implementations \cite{PhysRevLett.127.110505}.
Nine codes were designed with code rates between 0.50 and 0.90 for $q=4$ ($q=8)$, corresponding to a supported QBER range between $0$ and $18\%$ ($24.7\%$). We used 100000 nodes with a maximum of 150 iterations for the MC-DE, the QBER was swept in 20 steps in a short range below the best possible threshold. In the Differential Evolution, population sizes between 15 and 50, a differential weight of 0.85, and a crossover probability of 0.7 were used. A sparsity of at most 10 nonzero coefficients in the polynomial was enforced, with the maximum node degree chosen as $d_{\text{v,max}}=40$. The sparsity allowed for reasonable optimization complexity, the maximum node degree was chosen to avoid numerical instability which we observed for higher values.

The results of the optimization can be found in Table \ref{tab:template} in form of the node degree distributions \eqref{lamda} and their performance according to Density Evolution by reporting their simulated threshold (Density Evolution threshold, DET). The efficiency was evaluated for the highest supported QBER, noted as the ensemble efficiency (EEff). The all-zero codeword assumption was used for the optimization and evaluation, which holds for the given scenario of a symmetric channel  \cite{4787626}. For all rates, the designed thresholds are close to the theoretical bound \eqref{upper_bound_ir}. LDPC codes with a length of $n=30000$ symbols were constructed using Progressive Edge Growth \cite{1377521}, and a log-FFT-SPA decoder was used to reconcile the messages. The simulated performance of the finite-size codes can be seen in Figure \ref{fig:ldpc} for a span of different QBER values, each data point being the mean of 100 samples. We used the blind reconciliation scheme for rate-adaption \cite{blind}. The mean number of decoding tries required for Bob to successfully reconcile is also shown. The valley pattern visible in the efficiency of the LDPC codes is due to the transition between codes of different rate, and a slight degradation in performance for high ratios of puncturing or shortening. The decoder used a maximum of 100 decoding iterations. As expected for finite-size codes, they do not reach the asymptotic ensemble threshold but show sub-optimal performance \cite{6987278}.

\begin{table*}[] 
\small
\caption{Degree distributions for 4- and 8-dimensional nonbinary LDPC codes. The threshold calculated by Density Evolution (DET), the corresponding ensemble efficiency (EEff) and the node degree distribution in edge view.}
\vspace{2mm}
\centering 
\setlength{\tabcolsep}{2pt}
\begin{tabular}{l | c c l } 
 & \multicolumn{3}{l}{\textbf{4-Dimensional}}\\ 

Rate &  DET & EEff &  Ensemble (edge view)\\
\midrule 

0.90 & 0.022 & 1.067 &
$\lambda(x) = 0.069x + 0.207x^2 + 0.051x^5 + 0.235x^6 + 0.070x^{14} + 0.225x^{17}$\\
&&&\hspace{0.9cm} $+ 0.139x^{28}$\\


0.85 &0.037 &  1.045 & $\lambda(x) = 0.094x + 0.210x^2 + 0.217x^{6} + 0.088x^{9} + 0.040x^{10} + 0.118x^{22}$\\
&&&\hspace{0.9cm}$+ 0.159x^{24} + 0.071x^{28}$\\

0.80 & 0.053& 1.044 &$\lambda(x) = 0.102x + 0.204x^2 + 0.116x^{5} + 0.088x^{7} + 0.174x^{14} + 0.107x^{26}$\\
&&&\hspace{0.9cm}$ + 0.205x^{27}$\\

0.75 & 0.069 & 1.053 & $\lambda(x) = 0.107x + 0.245x^3 + 0.192x^{6} + 0.034x^{9} + 0.207x^{18} + 0.161x^{25}$\\
&&&\hspace{0.9cm}$+ 0.049x^{27}$\\

0.70 & 0.08 & 1.054 & $\lambda(x) = 0.113x + 0.245x^2 + 0.143x^4 + 0.081x^{10} + 0.066^{14} + 0.147^{16}$\\
&&&\hspace{0.9cm}$ + 0.034x^{23} + 0.168x^{24}$\\

0.65 & 0.11 & 1.045 & $\lambda(x) = 0.133x + 0.213x^2 + 0.207x^{5} + 0.014x^{8} + 0.022x^{17} + 0.171x^{19}$\\
&&&\hspace{0.9cm}$+ 0.237x^{27}$\\

0.60 & 0.13 & 1.047 & $\lambda(x) = 0.172x + 0.252x^2 + 0.216x^6 + 0.022x^{11} + + 0.075x^{12} + 0.077x^{18}$\\
&&&\hspace{0.9cm}$+ 0.183x^{20}$\\

0.55 & 0.15 & 1.041 & $\lambda(x) = 0.177x + 0.279x^2 + 0.147x^7 + 0.035x^8 + + 0.088x^{10} + 0.129x^{14}$\\
&&&\hspace{0.9cm}$ + 0.143x^{25}$\\

0.50 & 0.18 & 1.037 & $\lambda(x) = 0.184x + 0.245x^2 + 0.087x^6 + 0.156x^7 + + 0.067x^{17} + 0.020^{21}$\\
&&&\hspace{0.9cm}$ +0.196^{25} + 0.041^{28}$\\

\midrule
\midrule
 & \multicolumn{3}{l}{\textbf{8-Dimensional}}\\
0.90 & 0.031 &  1.060  & $\lambda(x) = 0.112x + 0.103x^2 + 0.194x^3 + 0.146x^9 + 0.163x^{10} + 0.003x^{17}$\\
&&&\hspace{0.9cm}$+ 0.173x^{19} + 0.049x^{26}  + 0.006x^{28} + 0.052x^{29}$\\

0.85 & 0.052 & 1.080 & $\lambda(x) = 0.125x + 0.165x^2 + 0.163x^5 + 0.11x^7 + 0.073x^{12} + 0.089x^{18}$ \\
&&&\hspace{0.9cm}$+ 0.122x^{27} + 0.154x^{32}$\\

0.80 & 0.072& 1.038 &$\lambda(x) = 0.146x + 0.177x^2 + 0.130x^4 + 0.084x^7 +  0.149x^{10} + 0.035x^{19}$\\
&&&\hspace{0.9cm}$+ 0.029x^{22} + 0.087x^{25} +  0.163x^{26}$\\

0.75 & 0.096 & 1.030 & $\lambda(x) = 0.165x + 0.192x^2 + 0.092x^5 + 0.176x^7 + 0.019x^{10} + 0.086x^{17}$\\
&&&\hspace{0.9cm}$ + 0.129x^{19} + 0.103x^{30} + 0.038x^{31}$\\

0.70 & 0.121 & 1.032 & $\lambda(x) = 0.160x + 0.208x^2 + 0.140x^5 + 0.096x^8 + 0.028x^{10} + 0.013x^{11}$\\
&&&\hspace{0.9cm}$+ 0.113x^{18} + 0.032x^{21} + 0.211x^{27}$\\

0.65 & 0.147 & 1.032 & $\lambda(x) = 0.173x + 0.228x^2 + 0.092x^4 + 0.169x^8 + 0.112x^{14} + 0.019x^{23}$\\
&&&\hspace{0.9cm}$+ 0.012x^{24} + 0.195x^{28}$\\

0.60 & 0.177 & 1.026 & $\lambda(x) = 0.192x + 0.196x^2 + 0.222x^5 + 0.104x^{13} + 0.114x^{23} + 0.055x^{25}$\\
&&&\hspace{0.9cm}$ + 0.117x^{27}$\\

0.55 & 0.207 & 1.024 & $\lambda(x) = 0.183x + 0.269x^2 + 0.124x^6 + 0.036x^8 + 0.097x^{10} + 0.004^{21}$\\
&&&\hspace{0.9cm}$ + 0.116x^{25} + 0.171x^{26}$\\

0.50 & 0.239 & 1.024 & $\lambda(x) = 0.215x + 0.256x^2 + 0.030x^4 + 0.154x^7 + 0.065x^{11} + 0.050^{13}$\\
&&&\hspace{0.9cm}$ + 0.072x^{21} + 0.128x^{27}$

\end{tabular}

\label{tab:template} 
\end{table*}

\begin{figure*}[h]
    \centering
    \includegraphics[width=\linewidth]{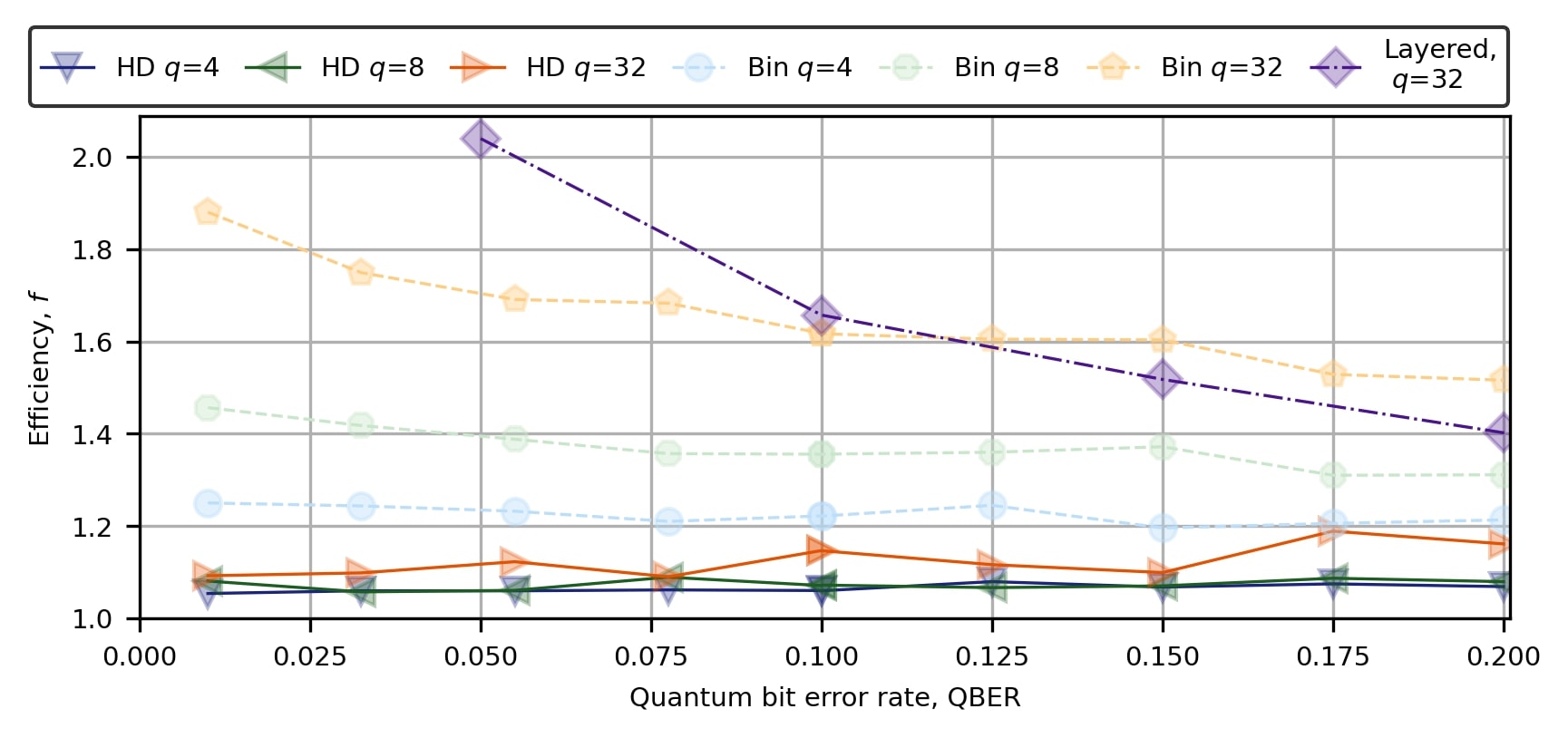}
    \caption{Efficiency of different approaches evaluated on a $q$-ary symmetric channel. Layered refers to the layered scheme, Bin to direct application of binary Cascade (serial), and HD to the high-dimensional Cascade proposed in this work. Data points represent the mean of 1000 samples and have a FER of less than $1\%$.}
    \label{fig:cascade}
\end{figure*}

\subsection{High-dimensional Cascade}

The performance of HD-Cascade, see Section \ref{hdc} was evaluated on the q-ary symmetric channel for dimensions $q = 4$, $8$, $32$, and for a QBER ranging from $1\%$ to $20\%$. The results are shown in Figure \ref{fig:cascade}. For comparison, a direct application of the best-performing Cascade modification on a binary mapping is also included. Binary Cascade's performance can be understood as generating an information leakage corresponding to an input key with an error rate of $\text{QBER}_{\text{BIN}}$. The proposed high-dimensional Cascade uses the same base Cascade with the additional adaptations discussed in Section \ref{hdc}. For $q=2$, HD-Cascade reduces to binary Cascade, resulting in equal performance. Both methods use the same block size of $n = 2^{16}$ bits for all cases. The used top-level block sizes $k_i$ for each iteration $i$ can be seen in Table \ref{tab:table1}, where $[\cdot]$ denotes rounding to the nearest integer. The block sizes have been chosen heuristically through numerical optimization. Additionally, the layered scheme is included as a reference \cite{6502993}. All data points have a frame error rate below 1\% and show an average of 1000 samples. The wave pattern observable for the efficiency of Cascade in Figure \ref{fig:cascade} and Figure \ref{fig:ldpc} is due to the integer rounding operation when calculating the block-sizes. Discretized block-sizes being a power of two have been shown to be optimal for the binary search in this setting \cite{pacher}.

\begin{table}[h]
   \centering
\caption{Block sizes used for HD-Cascade.} \label{tab:table1}
\def\arraystretch{1.25}%
\begin{tabular}{|c|c|c|c|c|c|c|c|}
         \hline \multicolumn{4}{|l|}{$k_1$} & \multicolumn{4}{l|}{$\min (2^{[\log_2(1/\text{QBER}_{\text{BIN}})]},n/2)$}\\
         \hline \multicolumn{4}{|l|}{$k_2$} &  \multicolumn{4}{l|}{$\min (2^{[\log_2(2q/\text{QBER}_{\text{BIN}})]},n/2)$}\\   
         \hline $k_3$ & $k_4$ & $k_5$ & $k_6$ & $n/16$ & $n/8$ & $n/4$ & $n/2$\\
         \hline
\end{tabular}
\end{table}%

\begin{figure*}
    \centering
    \includegraphics[width=\linewidth]{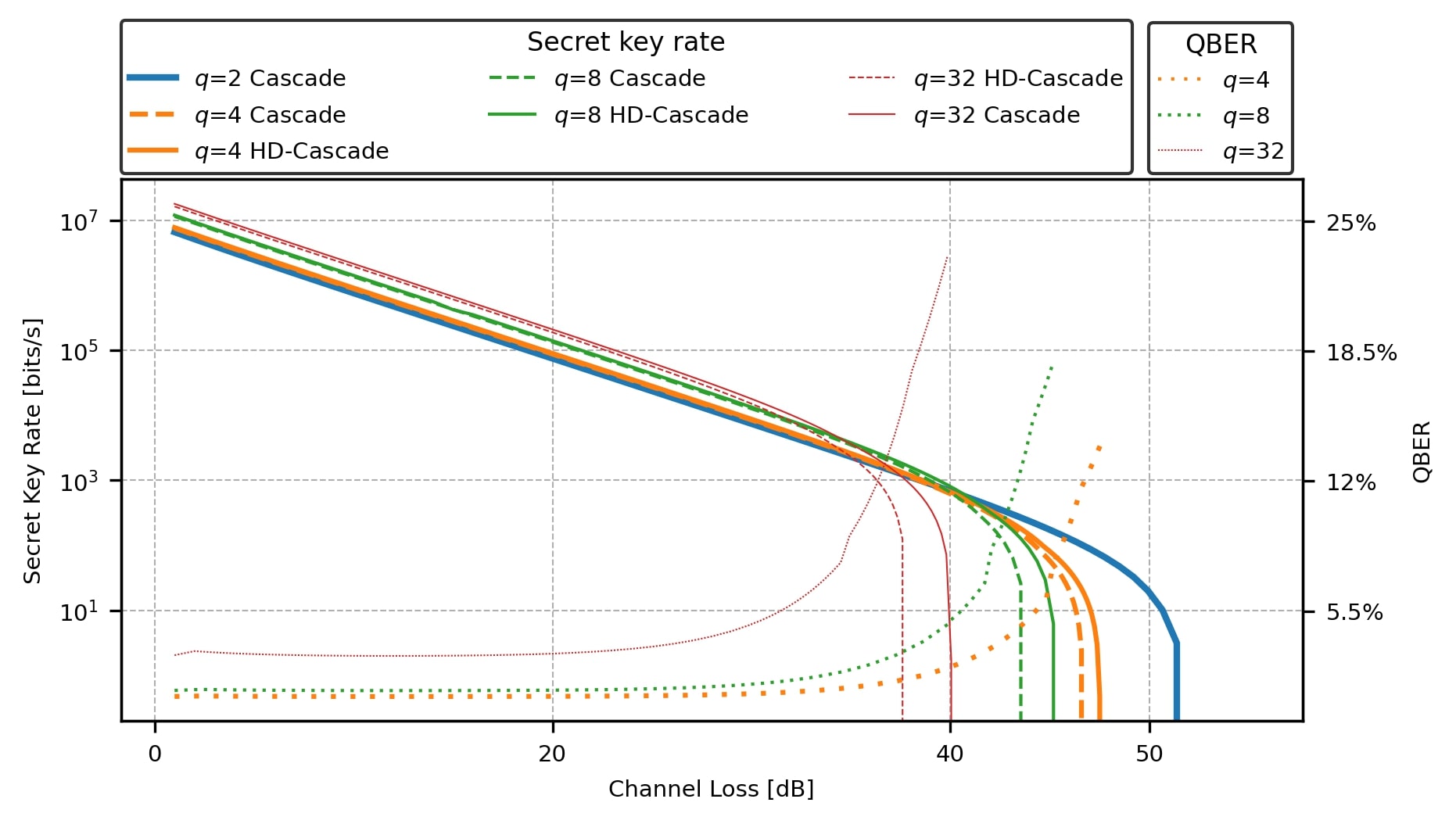}
    \caption{Secret key rate vs. channel loss for different dimensions. The decreasing dotted/dashed lines show results for direct application of binary Cascade whereas the solid lines show the performance for HD-Cascade. The increasing dotted/dashed lines show the respective QBER.}
    \label{fig:skr}
\end{figure*}

\noindent The increase in both the range and secret key rate resulting from using HD-Cascade instead of directly applying binary Cascade is depicted in Figure~\ref{fig:skr}. The used protocols are 1-decoy state QKD protocols \cite{Rusca_2018, Vagniluca_2020}, with the secret key length $l_q$ per block given as 

\begin{equation}
    l_q \leq \log_2(q) D_0^Z + D_1^Z (\log_2(q) - \text{H}_\text{HD}(\Phi_Z,q) ) - \text{leak}_{\text{IR}} - 6 \log_2(19/\epsilon_{\text{sec}}) - \log_2(2/\epsilon_{\text{cor}}),
\end{equation}

where $D_0^Z$ is a lower bound on vacuum events and $D_1^Z$ is a lower bound on single photon events. We refer to the supplementary information of \cite{Rusca_2018} for derivation of these bounds. $\text{H}_\text{HD}(\Phi_Z,q)$ is the high-dimensional Shannon entropy,

\begin{equation}
    \text{H}_\text{HD}(\Phi_Z,q) = -\Phi_Z \log_2(\Phi_Z / (q-1)) - (1-\Phi_Z) \log_2 (1-\Phi_Z),
\end{equation}

where $\Phi_Z$ is an upper bound on the phase error rate, $\epsilon_{\text{sec}}$ is the security parameter, and $\epsilon_{\text{cor}}$ is the correctness parameter. Experimental parameters for the simulation are derived from \cite{muji} for $q=2$ and $4$, where a combination of polarization and path is used to encode the qudits. For $q=8$ and $32$, we used a generalization of the setup. Additional losses might transpire due to increased experimental complexity which are not considered in the simulation. Some of the parameters are listed in Table \ref{TAB:paras}.

\begin{table}
    
    \centering
    \renewcommand{\arraystretch}{1.5} 
    \setlength{\arrayrulewidth}{1.5pt} 

    \begin{tabular}{|c|c|c|c|c|}
        \hline
        \cellcolor{gray!20}{\textcolor{black}{$\epsilon_{\text{sec}}$}} & \cellcolor{gray!20}{\textcolor{black}{Block-size privacy amplification}} & \cellcolor{gray!20}{\textcolor{black}{Repetition rate}} & \cellcolor{gray!20}{\textcolor{black}{Dead time}} & \cellcolor{gray!20}{\textcolor{black}{Insertion loss Z}} \\
        \hline
        $10^{-12}$ & $10^8$ & 487 MHz & 25 ns & 1 dB \\
        \hline
        \cellcolor{gray!20}{\textcolor{black}{$\epsilon_{\text{cor}}$}} & \cellcolor{gray!20}{\textcolor{black}{Misalignment error}} & \cellcolor{gray!20}{\textcolor{black}{Detector efficiency}} & \cellcolor{gray!20}{\textcolor{black}{Dark counts}} & \cellcolor{gray!20}{\textcolor{black}{Insertion loss X}} \\
        \hline
        $10^{-12}$ & 1$\%$ & $86\%$ & 330 & 3 dB \\
        \hline
    \end{tabular}
    \caption{Parameters used for the SKR simulations. Repetition rate refers to the repetition rate of the source, dead time to the dead time of the detectors used, the insertion loss refers to the loss of the receiver side for the two bases used, X and Z. The dark counts and detector efficiency are given for a single detector. For more details see \cite{muji, Rusca_2018, Vagniluca_2020}.}
     \label{TAB:paras}
\end{table}

The improvement in the relative secret key rate $r$ obtained using HD-Cascade is shown in Figure~\ref{fig:improv}. 
We also analysed the performance of HD-Cascade on experimental data provided by the experiment in \cite{muji}, which confirms the simulated performance.\\

\begin{figure}
    \centering
    \includegraphics[width=0.8\linewidth]{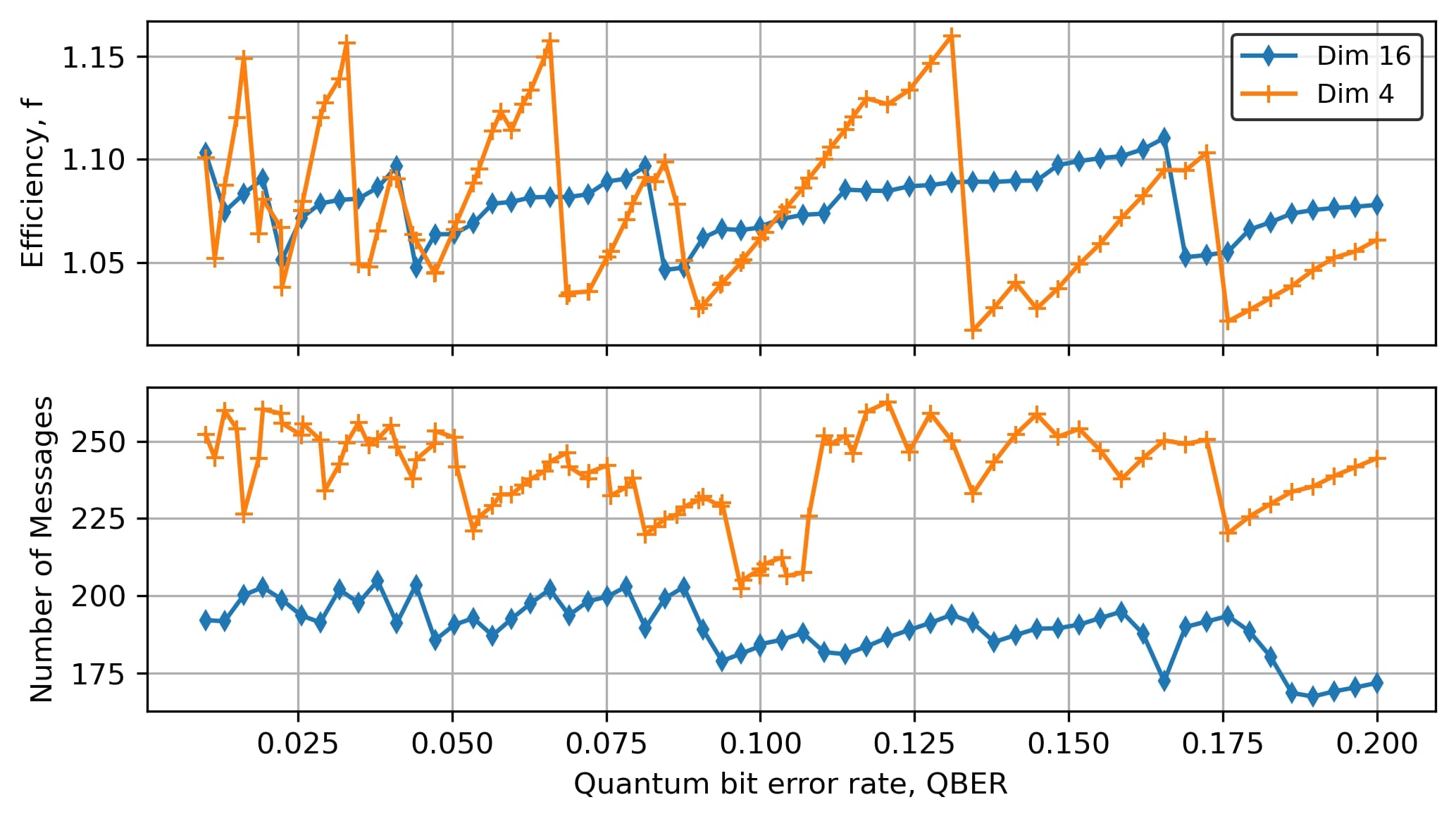}
    \caption{Top: The efficiency of the parallel high-dimensional Cascade implementation for dimensions 4 and 16. Bottom: The number of messages sent from Alice to Bob for a single frame.}
    \label{fig:prac}
\end{figure}

\begin{figure}
\begin{minipage}[c]{.49\textwidth}
  \centering
  \includegraphics[width=1\linewidth]{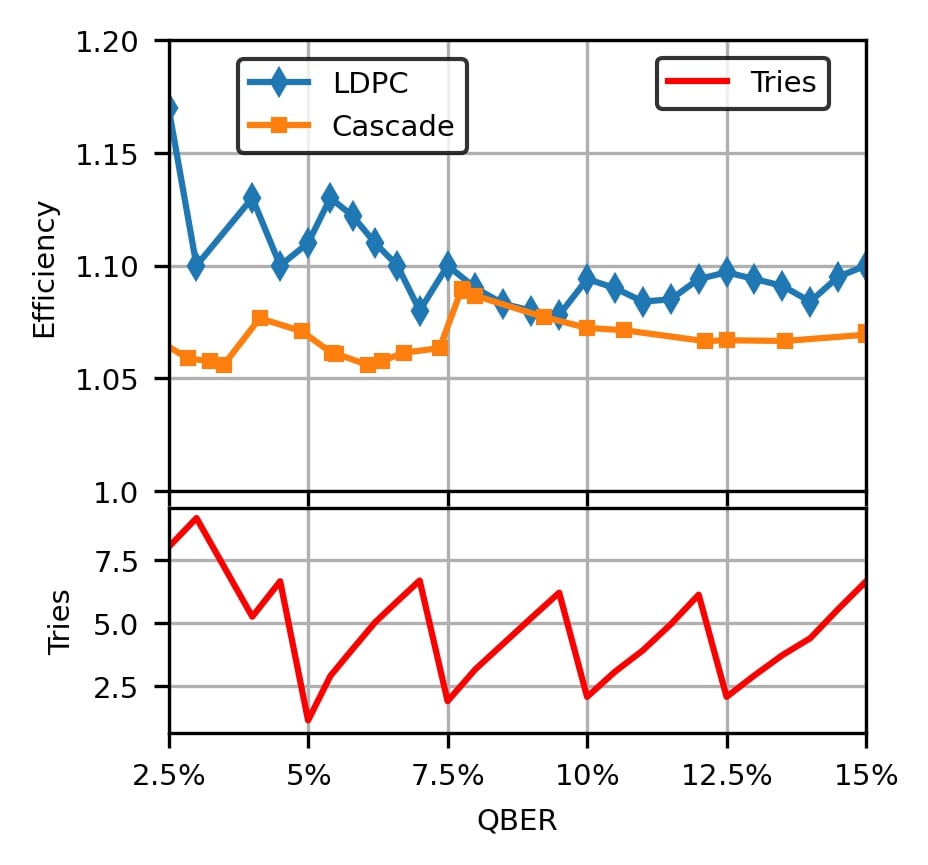}
  \caption{Top: Efficiency of using nonbinary LDPC codes and HD-Cascade for different QBER values for $q=8$. Bottom: Number of decoding tries in the blind scheme used for the LDPC codes, i.e. the number of messages required. The number of messages required for HD-Cascade can be seen in Figure \ref{fig:prac}.}
  \label{fig:ldpc}
\end{minipage}%
\hfill
\begin{minipage}[c]{.49\textwidth}
  \centering
  \includegraphics[width=1\linewidth, trim={0 0 0 1.1cm},clip]{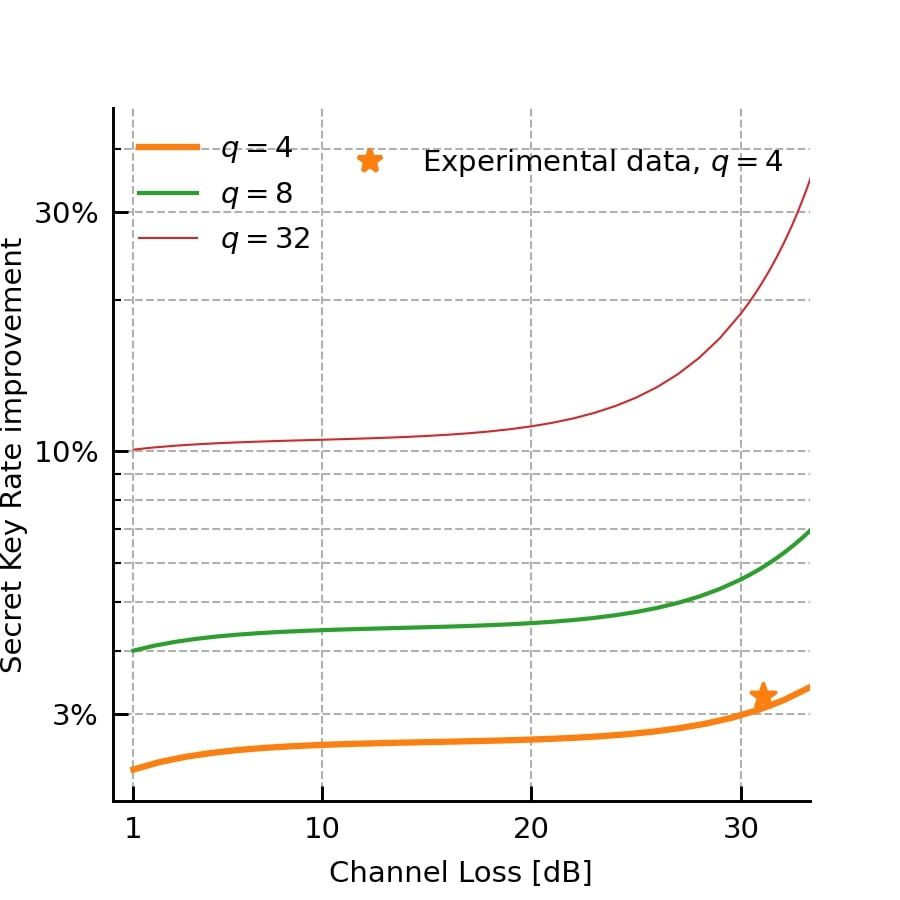}
  \caption{Relative improvement of the secret key rate for using HD-Cascade compared to binary Cascade. Experimental data provided by recent experiment\cite{muji}.}
  \label{fig:improv}
\end{minipage}
\end{figure}

\noindent We further analyse the performance of the parallel high-dimensional Cascade implementation for $q = 4, 16$. The results can be seen in Figure \ref{fig:prac}. While a slight penalty in efficiency can be observed, the number of messages sent for a single frame is greatly reduced compared to the serial approach. In the serial approach, the number of messages is roughly given by $n\text{H}(X|Y)$, i.e. around 8000 messages for $q=16$, QBER$=5\%$, and $f=1.05$, compared to a mean of 239 and 189 messages for $q=4$ and $q=16$ for the parallel implementation. Notably, the number of messages seems to decrease for higher dimensions. All data points represent the mean of 2000 samples, the frame error rate is below 0.1\% for all QBER values.

\section{Discussion}\label{sec:dis}

\subsection{Nonbinary LDPC codes}

Nonbinary LDPC codes are a natural candidate for the information reconciliation stage of HD-QKD, as their order can be matched to the dimension of the used qudits, and they are known to have good decoding performance \cite{706440}.  Although they typically come with increased decoding complexity, this drawback is less of a concern in this context, since the keys can be processed and stored before being employed in real-time applications, which reduces the significance of decoding latency. Nevertheless, less complex decoder algorithms like EMS \cite{4155118} or TEMS \cite{6125307} can be considered to allow the usage of longer codes and for increasing the throughput with a small penalty on efficiency.

\noindent The node degree distributions we constructed show ensemble efficiencies close to one, $1.037 - 1.067$ for $q=4$ and $1.024-1.080$ for $q=8$. To the best of our knowledge, there is no inherent reason for the efficiencies of $q=8$ to be lower than for $q=4$. It is rather just a heuristic result due to optimization parameters fitting better. Although the ensembles we found display thresholds near the Slepian-Wolf bound, we believe that even better results could be achieved by expanding the search of the hyperparameters involved in the optimization, such as the enforced sparsity and the highest degree of $\lambda$, and by performing a finer sweep of the QBER during Density Evolution.  The evaluated efficiency of finite-size codes shows them performing significantly worse than the thresholds computed with Density Evolution, with efficiencies ranging from $1.078$ to $1.14$ for QBER values in a medium range. This gap can be reduced by using longer codes and improving the code construction, e.g. using improved versions of the PEG algorithm \cite{IPEG1, IPEG2}. The dependency of the efficiency on the QBER can further be reduced, i.e. flatting the curve in Figure \ref{fig:ldpc},  by improving the position of punctured bits \cite{5649264}.

While working on this manuscript, the usage of nonbinary LDPC codes for information reconciliation has also been proposed in \cite{mitra2023nonbinary}. They suggest mapping symbols of high dimensionality to symbols of lower dimensionality but still higher than 2 if beneficial, in similarity to the layered scheme. This can further be used to decrease computational complexity if required.

\subsection{HD-Cascade}

HD-Cascade has improved performance on high-dimensional QKD setups compared to directly applying binary Cascade. We can see significant improvement in efficiency, with mean efficiencies of $f_{\text{HD-Cascade}}=1.06$, $1.07$, $1.12$ compared to $f_{\text{Cascade}}=1.22$, $1.36$, $1.65$ for $q=4$, $8$, $32$, respectively. Using the setup parameters of a recent experimental implementation of 4-dimensional QKD \cite{muji}, a resulting improvement in range and secret key rate can be observed, especially for higher dimensions. For $q=32$, an increase of more than $10\%$ in secret key rate over all QBER values and an additional $2.5$ dB in tolerable channel loss is achievable according to our simulation results.

The serial approach demonstrates high efficiency across all QBER values but suffers from a strong increase in execution time with higher error rates and an impractical requirement on communication rounds. Apart from the inherent scaling of Cascade with the QBER that is also present for binary implementations, this is additionally attributable to the immediate cascading of partner bits. This penalty can be greatly reduced when implementing the parallel high-dimensional version. The resulting penalty on the mean efficiency is very small with mean efficiencies of $f_{\text{HD-Cascade, practical}}=1.07$, $1.08$ for $q=4$, $16$, respectively. We again see the sawtooth pattern that can also be observed in binary Cascade and is attributed to discrete jumps in block sizes \cite{martinezmateo2014demystifying}. We observed a significantly higher variance in the efficiency with respect to QBER for the parallel approach, especially for $q=4$.  We assume that this is due to better-performing block size selection for $q=16$ and that an optimized parameter selection will reduce the sawtooth pattern and increase efficiency overall for all values of $q$. The search for optimized parameters has been investigated and is with great impact on the efficiency of binary Cascade, see again \cite{martinezmateo2014demystifying} for an overview. The search for well-performing block size selections for HD-Cascade could therefore be an interesting object of further research.

While the many rounds of communication required by Cascade have raised concerns about resulting limitations on throughput, recent research has shown that sophisticated software implementations can enable Cascade to achieve high throughput even with realistic latency on the classical channel \cite{mao2021high, pedersen2013high}. While high-dimensional Cascade has little additional computations compared to binary Cascade, the number of messages is increased due to the additional rounds of cascading on the partner-bits. For $q=4$, $16$, the number of messages per frame can be seen in Figure \ref{fig:prac}. Notably, the number of messages required seems to decrease with increasing dimension, resulting in an increasing throughput with increasing dimension. The impact on throughput is dependent on the block size, communication delay, and computing time, an analysis of this for the binary case can be found in \cite{pedersen2013high} and is required for the high-dimensional case before a general behaviour can be attributed. This is in contrast to LDPC codes whose throughput would scale negatively with increasing dimension. In our simulations, the relative increase in throughput between $q=4$ and $q=16$ is $30\%$ for a QBER of $1\%$ and a delay of 1 ms between Alice and Bob.


\subsection{Comparison}

Overall, HD-Cascade and nonbinary LDPC codes show good efficiency over all relevant QBER values, with HD-Cascade performing slightly better in terms of efficiency (see Figure~\ref{fig:ldpc}) at the expense of increased interactivity. Both show significant improvement compared to the layered scheme. The performance of the layered scheme can be seen for $q=32$ in Figure \ref{fig:skr}, notably for a much smaller block length (data read off Figure 5 in \cite{6502993}). Later experimental implementations report efficiencies of $f=1.25$ \cite{Liu2022HighdimensionalQK} ($q=3$, $n=1944$, $p=8\%$) and $f=1.17$ \cite{2015NJPh...17b2002Z} ($q=1024$, $n=4000$, $p=39.6\%)$. These papers report their efficiencies in the $\beta$-notation, we converted them to the $f$-notation for comparison. $\beta$ is commonly used in the continuous-variable QKD community, whereas $f$ is more widespread with respect to discrete-variable QKD. They can be related via \cite{martinezmateo2014demystifying}

\begin{equation}    
    \beta(\text{H}(X)-\text{H}(X|Y)) = \text{H}(X)-f\text{H}(X|Y).
\end{equation}\

\noindent HD-Cascade shows a flat efficiency behavior over all ranges, compared to the LDPC codes, which have a bad performance for very low QBER values and  show an increase in performance with increasing QBER, see Table \ref{tab:template}. This behavior can also be observed in LDPC codes used in binary QKD \cite{5649550, 5205475, Mao2019}. While the focus of this work lies in introducing new methods for high-dimensional information reconciliation with good efficiencies, the throughput is another important measure, especially with continuously improving input rates from advancing QKD hardware implementation. While an absolute and direct comparison of throughput strongly depends on the specific implementation and setup parameters and is not a contribution of this work, relative performances can be considered. Cascade has low computational complexity but high interactivity which can limit throughput in scenarios where the classical channel has a high latency.  

Nonbinary LDPC codes, on the other hand, have low requirements on interactivity (usually below 10 syndromes per frame using the blind scheme, see Figure \ref{fig:ldpc}) but high computational costs at the decoder. Their decoding complexity scales strongly with $q$ but only slightly with the QBER, as its main dependence is on the number of entries in its parity check matrix and the node degrees. It should be noted that the QBER is usually fairly stable until the loss approaches the maximum range of the setup, e.g. see Figure~\ref{fig:skr}, and that higher dimensions tend to operate at higher QBER values \cite{Cozzolino_2019}. It should be emphasized that for QKD, low latency in post-processing is often not a priority as keys do not need to be available immediately but can be stored for usage, with varying importance for different network scenarios. QKD systems can be significantly bigger and more expensive than setups for classical communication. This allows for reconciliation schemes with comparatively high latency and high computational complexity, for example by extensive usage of pipelining \cite{yang2021fpga, mao2021high, 6365815}.

\section{Conclusion} \label{sec:con}
We introduced two new methods for the information reconciliation stage of high dimensional Quantum Key Distribution and investigated their performance. We present nonbinary LDPC codes designed and optimized specifically for the use in high-dimensional QKD systems. They allow for reconciliation with good efficiency while maintaining a low interactivity between the two parties. For HD-QKD systems of dimension 8, the codes we constructed show efficiencies between $f=1.078$ and $f=1.14$ for QBER values between $3\%$ and $15\%$. We further propose new modifications to Cascade to make it suitable for high-dimensional QKD that greatly increase its efficiency on those systems. The main modification manifests in the request of partner-bits as a additional means of detecting errors in Cascade. Our simulations show mean efficiencies of $f=1.06$, $1.07$, and $1.12$ for dimensions $4$, $8$ and $32$. We also analyse the impact of HD-Cascade on the secret key rate compared to using binary Cascade and note significant improvement, i.e. more than $10\%$ for a 32-dimensional system for all possible channel losses and an increase in range corresponding to an additional 2.5 dB loss.

\section*{Declarations}

\textbf{Acknowledgements} The Center of Excellence SPOC (ref DNRF123).\\

\noindent \textbf{Conflict of interest} The authors declare no competing financial or non-financial interests.\\

\noindent\textbf{Data Availability} All data  used in this work are available from the corresponding author upon reasonable request.


\bibliography{sn-bibliography}


\begin{thebibliography}{62}
\ifx \bisbn   \undefined \def \bisbn  #1{ISBN #1}\fi
\ifx \binits  \undefined \def \binits#1{#1}\fi
\ifx \bauthor  \undefined \def \bauthor#1{#1}\fi
\ifx \batitle  \undefined \def \batitle#1{#1}\fi
\ifx \bjtitle  \undefined \def \bjtitle#1{#1}\fi
\ifx \bvolume  \undefined \def \bvolume#1{\textbf{#1}}\fi
\ifx \byear  \undefined \def \byear#1{#1}\fi
\ifx \bissue  \undefined \def \bissue#1{#1}\fi
\ifx \bfpage  \undefined \def \bfpage#1{#1}\fi
\ifx \blpage  \undefined \def \blpage #1{#1}\fi
\ifx \burl  \undefined \def \burl#1{\textsf{#1}}\fi
\ifx \doiurl  \undefined \def \doiurl#1{\url{https://doi.org/#1}}\fi
\ifx \betal  \undefined \def \betal{\textit{et al.}}\fi
\ifx \binstitute  \undefined \def \binstitute#1{#1}\fi
\ifx \binstitutionaled  \undefined \def \binstitutionaled#1{#1}\fi
\ifx \bctitle  \undefined \def \bctitle#1{#1}\fi
\ifx \beditor  \undefined \def \beditor#1{#1}\fi
\ifx \bpublisher  \undefined \def \bpublisher#1{#1}\fi
\ifx \bbtitle  \undefined \def \bbtitle#1{#1}\fi
\ifx \bedition  \undefined \def \bedition#1{#1}\fi
\ifx \bseriesno  \undefined \def \bseriesno#1{#1}\fi
\ifx \blocation  \undefined \def \blocation#1{#1}\fi
\ifx \bsertitle  \undefined \def \bsertitle#1{#1}\fi
\ifx \bsnm \undefined \def \bsnm#1{#1}\fi
\ifx \bsuffix \undefined \def \bsuffix#1{#1}\fi
\ifx \bparticle \undefined \def \bparticle#1{#1}\fi
\ifx \barticle \undefined \def \barticle#1{#1}\fi
\bibcommenthead
\ifx \bconfdate \undefined \def \bconfdate #1{#1}\fi
\ifx \botherref \undefined \def \botherref #1{#1}\fi
\ifx \url \undefined \def \url#1{\textsf{#1}}\fi
\ifx \bchapter \undefined \def \bchapter#1{#1}\fi
\ifx \bbook \undefined \def \bbook#1{#1}\fi
\ifx \bcomment \undefined \def \bcomment#1{#1}\fi
\ifx \oauthor \undefined \def \oauthor#1{#1}\fi
\ifx \citeauthoryear \undefined \def \citeauthoryear#1{#1}\fi
\ifx \endbibitem  \undefined \def \endbibitem {}\fi
\ifx \bconflocation  \undefined \def \bconflocation#1{#1}\fi
\ifx \arxivurl  \undefined \def \arxivurl#1{\textsf{#1}}\fi
\csname PreBibitemsHook\endcsname

\bibitem[\protect\citeauthoryear{Bennett and Brassard}{1984}]{BEN84}
\begin{bchapter}
\bauthor{\bsnm{Bennett}, \binits{C.H.}},
\bauthor{\bsnm{Brassard}, \binits{G.}}:
\bctitle{{Quantum cryptography: Public key distribution and coin tossing}}.
In: \bbtitle{Proceedings of {IEEE} International Conference on Computers, Systems, and Signal Processing},
\bconflocation{India},
p. \bfpage{175}
(\byear{1984})
\end{bchapter}
\endbibitem

\bibitem[\protect\citeauthoryear{Pirandola et~al.}{2020}]{Pirandola_2020}
\begin{barticle}
\bauthor{\bsnm{Pirandola}, \binits{S.}},
\bauthor{\bsnm{Andersen}, \binits{U.L.}},
\bauthor{\bsnm{Banchi}, \binits{L.}},
\bauthor{\bsnm{Berta}, \binits{M.}},
\bauthor{\bsnm{Bunandar}, \binits{D.}},
\bauthor{\bsnm{Colbeck}, \binits{R.}},
\bauthor{\bsnm{Englund}, \binits{D.}},
\bauthor{\bsnm{Gehring}, \binits{T.}},
\bauthor{\bsnm{Lupo}, \binits{C.}},
\bauthor{\bsnm{Ottaviani}, \binits{C.}},
\bauthor{\bsnm{Pereira}, \binits{J.L.}},
\bauthor{\bsnm{Razavi}, \binits{M.}},
\bauthor{\bsnm{Shaari}, \binits{J.S.}},
\bauthor{\bsnm{Tomamichel}, \binits{M.}},
\bauthor{\bsnm{Usenko}, \binits{V.C.}},
\bauthor{\bsnm{Vallone}, \binits{G.}},
\bauthor{\bsnm{Villoresi}, \binits{P.}},
\bauthor{\bsnm{Wallden}, \binits{P.}}:
\batitle{Advances in quantum cryptography}.
\bjtitle{Advances in Optics and Photonics}
\bvolume{12}(\bissue{4}),
\bfpage{1012}
(\byear{2020})
\doiurl{10.1364/aop.361502}
\end{barticle}
\endbibitem

\bibitem[\protect\citeauthoryear{Cozzolino et~al.}{2019}]{201900038}
\begin{barticle}
\bauthor{\bsnm{Cozzolino}, \binits{D.}},
\bauthor{\bsnm{Da~Lio}, \binits{B.}},
\bauthor{\bsnm{Bacco}, \binits{D.}},
\bauthor{\bsnm{Oxenl\o~we}, \binits{L.K.}}:
\batitle{High-dimensional quantum communication: Benefits, progress, and future challenges}.
\bjtitle{Advanced Quantum Technologies}
\bvolume{2}(\bissue{12}),
\bfpage{1900038}
(\byear{2019})
\doiurl{10.1002/qute.201900038}
\end{barticle}
\endbibitem

\bibitem[\protect\citeauthoryear{Bechmann-Pasquinucci and Tittel}{2000}]{PhysRevA.61.062308}
\begin{barticle}
\bauthor{\bsnm{Bechmann-Pasquinucci}, \binits{H.}},
\bauthor{\bsnm{Tittel}, \binits{W.}}:
\batitle{Quantum cryptography using larger alphabets}.
\bjtitle{Phys. Rev. A}
\bvolume{61},
\bfpage{062308}
(\byear{2000})
\doiurl{10.1103/PhysRevA.61.062308}
\end{barticle}
\endbibitem

\bibitem[\protect\citeauthoryear{Niu et~al.}{2016}]{PhysRevA.94.052323}
\begin{barticle}
\bauthor{\bsnm{Niu}, \binits{M.Y.}},
\bauthor{\bsnm{Xu}, \binits{F.}},
\bauthor{\bsnm{Shapiro}, \binits{J.H.}},
\bauthor{\bsnm{Furrer}, \binits{F.}}:
\batitle{Finite-key analysis for time-energy high-dimensional quantum key distribution}.
\bjtitle{Phys. Rev. A}
\bvolume{94},
\bfpage{052323}
(\byear{2016})
\doiurl{10.1103/PhysRevA.94.052323}
\end{barticle}
\endbibitem

\bibitem[\protect\citeauthoryear{Ecker et~al.}{2019}]{PhysRevX.9.041042}
\begin{barticle}
\bauthor{\bsnm{Ecker}, \binits{S.}},
\bauthor{\bsnm{Bouchard}, \binits{F.}},
\bauthor{\bsnm{Bulla}, \binits{L.}},
\bauthor{\bsnm{Brandt}, \binits{F.}},
\bauthor{\bsnm{Kohout}, \binits{O.}},
\bauthor{\bsnm{Steinlechner}, \binits{F.}},
\bauthor{\bsnm{Fickler}, \binits{R.}},
\bauthor{\bsnm{Malik}, \binits{M.}},
\bauthor{\bsnm{Guryanova}, \binits{Y.}},
\bauthor{\bsnm{Ursin}, \binits{R.}},
\bauthor{\bsnm{Huber}, \binits{M.}}:
\batitle{Overcoming noise in entanglement distribution}.
\bjtitle{Phys. Rev. X}
\bvolume{9},
\bfpage{041042}
(\byear{2019})
\doiurl{10.1103/PhysRevX.9.041042}
\end{barticle}
\endbibitem

\bibitem[\protect\citeauthoryear{Cozzolino et~al.}{2019}]{Cozzolino_2019}
\begin{barticle}
\bauthor{\bsnm{Cozzolino}, \binits{D.}},
\bauthor{\bsnm{Lio}, \binits{B.D.}},
\bauthor{\bsnm{Bacco}, \binits{D.}},
\bauthor{\bsnm{Oxenl{\o}we}, \binits{L.K.}}:
\batitle{High-dimensional quantum communication: Benefits, progress, and future challenges}.
\bjtitle{Advanced Quantum Technologies}
\bvolume{2}(\bissue{12}),
\bfpage{1900038}
(\byear{2019})
\doiurl{10.1002/qute.201900038}
\end{barticle}
\endbibitem

\bibitem[\protect\citeauthoryear{Zhou et~al.}{2013}]{6502993}
\begin{bchapter}
\bauthor{\bsnm{Zhou}, \binits{H.}},
\bauthor{\bsnm{Wang}, \binits{L.}},
\bauthor{\bsnm{Wornell}, \binits{G.}}:
\bctitle{Layered schemes for large-alphabet secret key distribution}.
In: \bbtitle{2013 Information Theory and Applications Workshop (ITA)},
pp. \bfpage{1}--\blpage{10}
(\byear{2013}).
\doiurl{10.1109/ITA.2013.6502993}
\end{bchapter}
\endbibitem

\bibitem[\protect\citeauthoryear{Davey and MacKay}{1998}]{706440}
\begin{bchapter}
\bauthor{\bsnm{Davey}, \binits{M.C.}},
\bauthor{\bsnm{MacKay}, \binits{D.J.C.}}:
\bctitle{Low density parity check codes over {GF(q)}}.
In: \bbtitle{1998 Information Theory Workshop (Cat. No.98EX131)},
pp. \bfpage{70}--\blpage{71}
(\byear{1998}).
\doiurl{10.1109/ITW.1998.706440}
\end{bchapter}
\endbibitem

\bibitem[\protect\citeauthoryear{Brassard and Salvail}{1994}]{10.1007/3-540-48285-7_35}
\begin{bchapter}
\bauthor{\bsnm{Brassard}, \binits{G.}},
\bauthor{\bsnm{Salvail}, \binits{L.}}:
\bctitle{Secret-key reconciliation by public discussion}.
In: \beditor{\bsnm{Helleseth}, \binits{T.}} (ed.)
\bbtitle{Advances in Cryptology --- EUROCRYPT '93},
pp. \bfpage{410}--\blpage{423}.
\bpublisher{Springer},
\blocation{Berlin, Heidelberg}
(\byear{1994})
\end{bchapter}
\endbibitem

\bibitem[\protect\citeauthoryear{Martinez-Mateo et~al.}{2012}]{blind}
\begin{barticle}
\bauthor{\bsnm{Martinez-Mateo}, \binits{J.}},
\bauthor{\bsnm{Elkouss}, \binits{D.}},
\bauthor{\bsnm{Martin}, \binits{V.}}:
\batitle{Blind reconciliation}.
\bjtitle{Quantum Info. Comput.}
\bvolume{12}(\bissue{9–10}),
\bfpage{791}--\blpage{812}
(\byear{2012})
\end{barticle}
\endbibitem

\bibitem[\protect\citeauthoryear{Kiktenko et~al.}{2021}]{Kiktenko_2021}
\begin{barticle}
\bauthor{\bsnm{Kiktenko}, \binits{E.O.}},
\bauthor{\bsnm{Malyshev}, \binits{A.O.}},
\bauthor{\bsnm{Fedorov}, \binits{A.K.}}:
\batitle{Blind information reconciliation with polar codes for quantum key distribution}.
\bjtitle{{IEEE} Communications Letters}
\bvolume{25}(\bissue{1}),
\bfpage{79}--\blpage{83}
(\byear{2021})
\doiurl{10.1109/lcomm.2020.3021142}
\end{barticle}
\endbibitem

\bibitem[\protect\citeauthoryear{Martinez-Mateo et~al.}{2013}]{Martinez-Mateo2013}
\begin{barticle}
\bauthor{\bsnm{Martinez-Mateo}, \binits{J.}},
\bauthor{\bsnm{Elkouss}, \binits{D.}},
\bauthor{\bsnm{Martin}, \binits{V.}}:
\batitle{Key reconciliation for high performance quantum key distribution}.
\bjtitle{Scientific Reports}
\bvolume{3}(\bissue{1}),
\bfpage{1576}
(\byear{2013})
\doiurl{10.1038/srep01576}
\end{barticle}
\endbibitem

\bibitem[\protect\citeauthoryear{Mao et~al.}{2022}]{mao2021high}
\begin{barticle}
\bauthor{\bsnm{Mao}, \binits{H.-K.}},
\bauthor{\bsnm{Li}, \binits{Q.}},
\bauthor{\bsnm{Hao}, \binits{P.-L.}},
\bauthor{\bsnm{Abd-El-Atty}, \binits{B.}},
\bauthor{\bsnm{Iliyasu}, \binits{A.M.}}:
\batitle{High performance reconciliation for practical quantum key distribution systems}.
\bjtitle{Optical and Quantum Electronics}
\bvolume{54}(\bissue{3}),
\bfpage{163}
(\byear{2022})
\doiurl{10.1007/s11082-021-03489-4}
\end{barticle}
\endbibitem

\bibitem[\protect\citeauthoryear{Pedersen and Toyran}{2015}]{pedersen2013high}
\begin{barticle}
\bauthor{\bsnm{Pedersen}, \binits{T.B.}},
\bauthor{\bsnm{Toyran}, \binits{M.}}:
\batitle{High performance information reconciliation for qkd with cascade}.
\bjtitle{Quantum Info. Comput.}
\bvolume{15}(\bissue{5–6}),
\bfpage{419}--\blpage{434}
(\byear{2015})
\end{barticle}
\endbibitem

\bibitem[\protect\citeauthoryear{Van~Assche et~al.}{2004}]{1266817}
\begin{barticle}
\bauthor{\bsnm{Van~Assche}, \binits{G.}},
\bauthor{\bsnm{Cardinal}, \binits{J.}},
\bauthor{\bsnm{Cerf}, \binits{N.J.}}:
\batitle{Reconciliation of a quantum-distributed gaussian key}.
\bjtitle{IEEE Transactions on Information Theory}
\bvolume{50}(\bissue{2}),
\bfpage{394}--\blpage{400}
(\byear{2004})
\doiurl{10.1109/TIT.2003.822618}
\end{barticle}
\endbibitem

\bibitem[\protect\citeauthoryear{Mao et~al.}{2019}]{Mao2019}
\begin{barticle}
\bauthor{\bsnm{Mao}, \binits{H.}},
\bauthor{\bsnm{Li}, \binits{Q.}},
\bauthor{\bsnm{Han}, \binits{Q.}},
\bauthor{\bsnm{Guo}, \binits{H.}}:
\batitle{High-throughput and low-cost {LDPC} reconciliation for quantum key distribution}.
\bjtitle{Quantum Information Processing}
\bvolume{18}(\bissue{7}),
\bfpage{232}
(\byear{2019})
\doiurl{10.1007/s11128-019-2342-2}
\end{barticle}
\endbibitem

\bibitem[\protect\citeauthoryear{Elkouss et~al.}{2009}]{5205475}
\begin{bchapter}
\bauthor{\bsnm{Elkouss}, \binits{D.}},
\bauthor{\bsnm{Leverrier}, \binits{A.}},
\bauthor{\bsnm{Alleaume}, \binits{R.}},
\bauthor{\bsnm{Boutros}, \binits{J.J.}}:
\bctitle{Efficient reconciliation protocol for discrete-variable quantum key distribution}.
In: \bbtitle{2009 IEEE International Symposium on Information Theory},
pp. \bfpage{1879}--\blpage{1883}
(\byear{2009}).
\doiurl{10.1109/ISIT.2009.5205475}
\end{bchapter}
\endbibitem

\bibitem[\protect\citeauthoryear{Kiktenko et~al.}{2017}]{PhysRevApplied.8.044017}
\begin{barticle}
\bauthor{\bsnm{Kiktenko}, \binits{E.O.}},
\bauthor{\bsnm{Trushechkin}, \binits{A.S.}},
\bauthor{\bsnm{Lim}, \binits{C.C.W.}},
\bauthor{\bsnm{Kurochkin}, \binits{Y.V.}},
\bauthor{\bsnm{Fedorov}, \binits{A.K.}}:
\batitle{Symmetric blind information reconciliation for quantum key distribution}.
\bjtitle{Phys. Rev. Appl.}
\bvolume{8},
\bfpage{044017}
(\byear{2017})
\doiurl{10.1103/PhysRevApplied.8.044017}
\end{barticle}
\endbibitem

\bibitem[\protect\citeauthoryear{Treeviriyanupab and Zhang}{2024}]{reviewer_bin_ir}
\begin{botherref}
\oauthor{\bsnm{Treeviriyanupab}, \binits{P.}},
\oauthor{\bsnm{Zhang}, \binits{C.-M.}}:
Efficient integration of rate-adaptive reconciliation with syndrome-based error estimation and subblock confirmation for quantum key distribution.
Entropy
\textbf{26}(1)
(2024)
\doiurl{10.3390/e26010053}
\end{botherref}
\endbibitem

\bibitem[\protect\citeauthoryear{Kasai et~al.}{2010}]{5649550}
\begin{bchapter}
\bauthor{\bsnm{Kasai}, \binits{K.}},
\bauthor{\bsnm{Matsumoto}, \binits{R.}},
\bauthor{\bsnm{Sakaniwa}, \binits{K.}}:
\bctitle{Information reconciliation for {QKD} with rate-compatible non-binary {LDPC} codes}.
In: \bbtitle{2010 International Symposium On Information Theory and Its Applications},
pp. \bfpage{922}--\blpage{927}
(\byear{2010}).
\doiurl{10.1109/ISITA.2010.5649550}
\end{bchapter}
\endbibitem

\bibitem[\protect\citeauthoryear{Pacher et~al.}{2015}]{pacher}
\begin{bchapter}
\bauthor{\bsnm{Pacher}, \binits{C.}},
\bauthor{\bsnm{Grabenweger}, \binits{P.}},
\bauthor{\bsnm{Martinez~Mateo}, \binits{J.}},
\bauthor{\bsnm{Martin}, \binits{V.}}:
\bctitle{An information reconciliation protocol for secret-key agreement with small leakage},
pp. \bfpage{730}--\blpage{734}
(\byear{2015}).
\doiurl{10.1109/ISIT.2015.7282551}
\end{bchapter}
\endbibitem

\bibitem[\protect\citeauthoryear{Liveris et~al.}{2002}]{1042242}
\begin{barticle}
\bauthor{\bsnm{Liveris}, \binits{A.D.}},
\bauthor{\bsnm{Xiong}, \binits{Z.}},
\bauthor{\bsnm{Georghiades}, \binits{C.N.}}:
\batitle{Compression of binary sources with side information at the decoder using {LDPC} codes}.
\bjtitle{IEEE Communications Letters}
\bvolume{6}(\bissue{10}),
\bfpage{440}--\blpage{442}
(\byear{2002})
\doiurl{10.1109/LCOMM.2002.804244}
\end{barticle}
\endbibitem

\bibitem[\protect\citeauthoryear{Bennett et~al.}{1988}]{pa_org}
\begin{barticle}
\bauthor{\bsnm{Bennett}, \binits{C.H.}},
\bauthor{\bsnm{Brassard}, \binits{G.}},
\bauthor{\bsnm{Robert}, \binits{J.-M.}}:
\batitle{Privacy amplification by public discussion}.
\bjtitle{SIAM Journal on Computing}
\bvolume{17}(\bissue{2}),
\bfpage{210}--\blpage{229}
(\byear{1988})
\doiurl{10.1137/0217014}
{\href{https://arxiv.org/abs/https://doi.org/10.1137/0217014}{{https://doi.org/10.1137/0217014}}}
\end{barticle}
\endbibitem

\bibitem[\protect\citeauthoryear{Lo}{2003}]{Lo_2003}
\begin{barticle}
\bauthor{\bsnm{Lo}, \binits{H.-K.}}:
\batitle{Method for decoupling error correction from privacy amplification}.
\bjtitle{New Journal of Physics}
\bvolume{5},
\bfpage{36}--\blpage{36}
(\byear{2003})
\doiurl{10.1088/1367-2630/5/1/336}
\end{barticle}
\endbibitem

\bibitem[\protect\citeauthoryear{Tupkary and L\"utkenhaus}{2023}]{tupkary2023using}
\begin{barticle}
\bauthor{\bsnm{Tupkary}, \binits{D.}},
\bauthor{\bsnm{L\"utkenhaus}, \binits{N.}}:
\batitle{Using cascade in quantum key distribution}.
\bjtitle{Phys. Rev. Appl.}
\bvolume{20},
\bfpage{064040}
(\byear{2023})
\doiurl{10.1103/PhysRevApplied.20.064040}
\end{barticle}
\endbibitem

\bibitem[\protect\citeauthoryear{Slepian and Wolf}{1973}]{1055037}
\begin{barticle}
\bauthor{\bsnm{Slepian}, \binits{D.}},
\bauthor{\bsnm{Wolf}, \binits{J.}}:
\batitle{Noiseless coding of correlated information sources}.
\bjtitle{IEEE Transactions on Information Theory}
\bvolume{19}(\bissue{4}),
\bfpage{471}--\blpage{480}
(\byear{1973})
\doiurl{10.1109/TIT.1973.1055037}
\end{barticle}
\endbibitem

\bibitem[\protect\citeauthoryear{Tomamichel et~al.}{2017}]{Tomamichel_2017}
\begin{botherref}
\oauthor{\bsnm{Tomamichel}, \binits{M.}},
\oauthor{\bsnm{Martinez-Mateo}, \binits{J.}},
\oauthor{\bsnm{Pacher}, \binits{C.}},
\oauthor{\bsnm{Elkouss}, \binits{D.}}:
Fundamental finite key limits for one-way information reconciliation in quantum key distribution.
Quantum Information Processing
\textbf{16}(11)
(2017)
\doiurl{10.1007/s11128-017-1709-5}
\end{botherref}
\endbibitem

\bibitem[\protect\citeauthoryear{Lütkenhaus}{1999}]{Luetkenhaus_1999}
\begin{barticle}
\bauthor{\bsnm{Lütkenhaus}, \binits{N.}}:
\batitle{Estimates for practical quantum cryptography}.
\bjtitle{Physical Review A}
\bvolume{59}(\bissue{5}),
\bfpage{3301}--\blpage{3319}
(\byear{1999})
\doiurl{10.1103/physreva.59.3301}
\end{barticle}
\endbibitem

\bibitem[\protect\citeauthoryear{Fung et~al.}{2010}]{Fung_2010}
\begin{botherref}
\oauthor{\bsnm{Fung}, \binits{C.-H.F.}},
\oauthor{\bsnm{Ma}, \binits{X.}},
\oauthor{\bsnm{Chau}, \binits{H.F.}}:
Practical issues in quantum-key-distribution postprocessing.
Physical Review A
\textbf{81}(1)
(2010)
\doiurl{10.1103/physreva.81.012318}
\end{botherref}
\endbibitem

\bibitem[\protect\citeauthoryear{Boaron et~al.}{2018}]{Boaron_2018}
\begin{barticle}
\bauthor{\bsnm{Boaron}, \binits{A.}},
\bauthor{\bsnm{Korzh}, \binits{B.}},
\bauthor{\bsnm{Houlmann}, \binits{R.}},
\bauthor{\bsnm{Boso}, \binits{G.}},
\bauthor{\bsnm{Rusca}, \binits{D.}},
\bauthor{\bsnm{Gray}, \binits{S.}},
\bauthor{\bsnm{Li}, \binits{M.-J.}},
\bauthor{\bsnm{Nolan}, \binits{D.}},
\bauthor{\bsnm{Martin}, \binits{A.}},
\bauthor{\bsnm{Zbinden}, \binits{H.}}:
\batitle{Simple 2.5{\hspace{0.167em}}{GHz} time-bin quantum key distribution}.
\bjtitle{Applied Physics Letters}
\bvolume{112}(\bissue{17}),
\bfpage{171108}
(\byear{2018})
\doiurl{10.1063/1.5027030}
\end{barticle}
\endbibitem

\bibitem[\protect\citeauthoryear{Lin and Li}{2021}]{ldpcbook}
\begin{bbook}
\bauthor{\bsnm{Lin}, \binits{S.}},
\bauthor{\bsnm{Li}, \binits{J.}}:
\bbtitle{Fundamentals of Classical and Modern Error-Correcting Codes}.
\bpublisher{Cambridge University Press},
\blocation{Cambridge}
(\byear{2021})
\end{bbook}
\endbibitem

\bibitem[\protect\citeauthoryear{Li et~al.}{2009}]{multis}
\begin{barticle}
\bauthor{\bsnm{Li}, \binits{G.}},
\bauthor{\bsnm{Fair}, \binits{I.}},
\bauthor{\bsnm{Krzymien}, \binits{W.}}:
\batitle{Density evolution for nonbinary ldpc codes under gaussian approximation}.
\bjtitle{Information Theory, IEEE Transactions on}
\bvolume{55},
\bfpage{997}--\blpage{1015}
(\byear{2009})
\doiurl{10.1109/TIT.2008.2011435}
\end{barticle}
\endbibitem

\bibitem[\protect\citeauthoryear{Aruna and Anbuselvi}{2013}]{6577118}
\begin{bchapter}
\bauthor{\bsnm{Aruna}, \binits{S.}},
\bauthor{\bsnm{Anbuselvi}, \binits{M.}}:
\bctitle{{FFT-SPA} based non-binary {LDPC} decoder for {IEEE} 802.11n standard}.
In: \bbtitle{2013 International Conference on Communication and Signal Processing},
pp. \bfpage{566}--\blpage{569}
(\byear{2013}).
\doiurl{10.1109/iccsp.2013.6577118}
\end{bchapter}
\endbibitem

\bibitem[\protect\citeauthoryear{Wymeersch et~al.}{2004}]{1312606}
\begin{bchapter}
\bauthor{\bsnm{Wymeersch}, \binits{H.}},
\bauthor{\bsnm{Steendam}, \binits{H.}},
\bauthor{\bsnm{Moeneclaey}, \binits{M.}}:
\bctitle{Log-domain decoding of {LDPC}codes over {GF}(q)}.
In: \bbtitle{2004 IEEE International Conference on Communications (IEEE Cat. No.04CH37577)},
vol. \bseriesno{2},
pp. \bfpage{772}--\blpage{7762}
(\byear{2004}).
\doiurl{10.1109/ICC.2004.1312606}
\end{bchapter}
\endbibitem

\bibitem[\protect\citeauthoryear{Li et~al.}{2009}]{4787626}
\begin{barticle}
\bauthor{\bsnm{Li}, \binits{G.}},
\bauthor{\bsnm{Fair}, \binits{I.J.}},
\bauthor{\bsnm{Krzymien}, \binits{W.A.}}:
\batitle{Density evolution for nonbinary {LDPC} codes under gaussian approximation}.
\bjtitle{IEEE Transactions on Information Theory}
\bvolume{55}(\bissue{3}),
\bfpage{997}--\blpage{1015}
(\byear{2009})
\doiurl{10.1109/TIT.2008.2011435}
\end{barticle}
\endbibitem

\bibitem[\protect\citeauthoryear{Dupraz et~al.}{2015}]{6987278}
\begin{barticle}
\bauthor{\bsnm{Dupraz}, \binits{E.}},
\bauthor{\bsnm{Savin}, \binits{V.}},
\bauthor{\bsnm{Kieffer}, \binits{M.}}:
\batitle{Density evolution for the design of non-binary low density parity check codes for {Slepian-Wolf} coding}.
\bjtitle{IEEE Transactions on Communications}
\bvolume{63}(\bissue{1}),
\bfpage{25}--\blpage{36}
(\byear{2015})
\doiurl{10.1109/TCOMM.2014.2382126}
\end{barticle}
\endbibitem

\bibitem[\protect\citeauthoryear{Elkouss et~al.}{2010a}]{Elkouss_2010_rate_adaptive}
\begin{bchapter}
\bauthor{\bsnm{Elkouss}, \binits{D.}},
\bauthor{\bsnm{Martinez-Mateo}, \binits{J.}},
\bauthor{\bsnm{Martin}, \binits{V.}}:
\bctitle{Secure rate-adaptive reconciliation}.
In: \bbtitle{2010 International Symposium On Information Theory and Its Applications}.
\bpublisher{IEEE}, \blocation{???}
(\byear{2010}).
\doiurl{10.1109/isita.2010.5650099} .
\burl{http://dx.doi.org/10.1109/ISITA.2010.5650099}
\end{bchapter}
\endbibitem

\bibitem[\protect\citeauthoryear{Elkouss et~al.}{2010b}]{5650099}
\begin{bchapter}
\bauthor{\bsnm{Elkouss}, \binits{D.}},
\bauthor{\bsnm{Martínez-Mateo}, \binits{J.}},
\bauthor{\bsnm{Martín}, \binits{V.}}:
\bctitle{Secure rate-adaptive reconciliation}.
In: \bbtitle{2010 International Symposium On Information Theory and Its Applications},
pp. \bfpage{179}--\blpage{184}
(\byear{2010}).
\doiurl{10.1109/ISITA.2010.5650099}
\end{bchapter}
\endbibitem

\bibitem[\protect\citeauthoryear{Kiktenko et~al.}{2017}]{symm_blind}
\begin{barticle}
\bauthor{\bsnm{Kiktenko}, \binits{E.O.}},
\bauthor{\bsnm{Trushechkin}, \binits{A.S.}},
\bauthor{\bsnm{Lim}, \binits{C.C.W.}},
\bauthor{\bsnm{Kurochkin}, \binits{Y.V.}},
\bauthor{\bsnm{Fedorov}, \binits{A.K.}}:
\batitle{Symmetric blind information reconciliation for quantum key distribution}.
\bjtitle{Phys. Rev. Appl.}
\bvolume{8},
\bfpage{044017}
(\byear{2017})
\doiurl{10.1103/PhysRevApplied.8.044017}
\end{barticle}
\endbibitem

\bibitem[\protect\citeauthoryear{Savin}{2008}]{4712621}
\begin{bchapter}
\bauthor{\bsnm{Savin}, \binits{V.}}:
\bctitle{Non binary {LDPC} codes over the binary erasure channel: Density evolution analysis}.
In: \bbtitle{2008 First International Symposium on Applied Sciences on Biomedical and Communication Technologies},
pp. \bfpage{1}--\blpage{5}
(\byear{2008}).
\doiurl{10.1109/ISABEL.2008.4712621}
\end{bchapter}
\endbibitem

\bibitem[\protect\citeauthoryear{Richardson and Urbanke}{2001}]{910577}
\begin{barticle}
\bauthor{\bsnm{Richardson}, \binits{T.J.}},
\bauthor{\bsnm{Urbanke}, \binits{R.L.}}:
\batitle{The capacity of low-density parity-check codes under message-passing decoding}.
\bjtitle{IEEE Transactions on Information Theory}
\bvolume{47}(\bissue{2}),
\bfpage{599}--\blpage{618}
(\byear{2001})
\doiurl{10.1109/18.910577}
\end{barticle}
\endbibitem

\bibitem[\protect\citeauthoryear{Gorgoglione et~al.}{2010}]{inproceedings}
\begin{bchapter}
\bauthor{\bsnm{Gorgoglione}, \binits{M.}},
\bauthor{\bsnm{Savin}, \binits{V.}},
\bauthor{\bsnm{Declercq}, \binits{D.}}:
\bctitle{Optimized puncturing distributions for irregular non-binary {LDPC} codes}.
In: \bbtitle{2010 International Symposium On Information Theory \& Its Applications},
pp. \bfpage{400}--\blpage{405}
(\byear{2010}).
\doiurl{10.1109/ISITA.2010.5649264}
\end{bchapter}
\endbibitem

\bibitem[\protect\citeauthoryear{Chung et~al.}{2001}]{910580}
\begin{barticle}
\bauthor{\bsnm{Chung}, \binits{S.-Y.}},
\bauthor{\bsnm{Richardson}, \binits{T.J.}},
\bauthor{\bsnm{Urbanke}, \binits{R.L.}}:
\batitle{Analysis of sum-product decoding of low-density parity-check codes using a gaussian approximation}.
\bjtitle{IEEE Transactions on Information Theory}
\bvolume{47}(\bissue{2}),
\bfpage{657}--\blpage{670}
(\byear{2001})
\doiurl{10.1109/18.910580}
\end{barticle}
\endbibitem

\bibitem[\protect\citeauthoryear{Storn and Price}{1997}]{de}
\begin{barticle}
\bauthor{\bsnm{Storn}, \binits{R.}},
\bauthor{\bsnm{Price}, \binits{K.}}:
\batitle{Differential evolution - a simple and efficient heuristic for global optimization over continuous spaces}.
\bjtitle{Journal of Global Optimization}
\bvolume{11},
\bfpage{341}--\blpage{359}
(\byear{1997})
\doiurl{10.1023/A:1008202821328}
\end{barticle}
\endbibitem

\bibitem[\protect\citeauthoryear{Buttler et~al.}{2003}]{winnow}
\begin{barticle}
\bauthor{\bsnm{Buttler}, \binits{W.T.}},
\bauthor{\bsnm{Lamoreaux}, \binits{S.K.}},
\bauthor{\bsnm{Torgerson}, \binits{J.R.}},
\bauthor{\bsnm{Nickel}, \binits{G.H.}},
\bauthor{\bsnm{Donahue}, \binits{C.H.}},
\bauthor{\bsnm{Peterson}, \binits{C.G.}}:
\batitle{Fast, efficient error reconciliation for quantum cryptography}.
\bjtitle{Phys. Rev. A}
\bvolume{67},
\bfpage{052303}
(\byear{2003})
\doiurl{10.1103/PhysRevA.67.052303}
\end{barticle}
\endbibitem

\bibitem[\protect\citeauthoryear{Martínez~Mateo et~al.}{2014}]{martinezmateo2014demystifying}
\begin{botherref}
\oauthor{\bsnm{Martínez~Mateo}, \binits{J.}},
\oauthor{\bsnm{Pacher}, \binits{C.}},
\oauthor{\bsnm{Peev}, \binits{M.}},
\oauthor{\bsnm{Ciurana}, \binits{A.}},
\oauthor{\bsnm{Martin}, \binits{V.}}:
Demystifying the information reconciliation protocol cascade.
Quantum Information and Computation
\textbf{15}
(2014)
\doiurl{10.26421/QIC15.5-6-6}
\end{botherref}
\endbibitem

\bibitem[\protect\citeauthoryear{Hu et~al.}{2021}]{PhysRevLett.127.110505}
\begin{barticle}
\bauthor{\bsnm{Hu}, \binits{X.-M.}},
\bauthor{\bsnm{Zhang}, \binits{C.}},
\bauthor{\bsnm{Guo}, \binits{Y.}},
\bauthor{\bsnm{Wang}, \binits{F.-X.}},
\bauthor{\bsnm{Xing}, \binits{W.-B.}},
\bauthor{\bsnm{Huang}, \binits{C.-X.}},
\bauthor{\bsnm{Liu}, \binits{B.-H.}},
\bauthor{\bsnm{Huang}, \binits{Y.-F.}},
\bauthor{\bsnm{Li}, \binits{C.-F.}},
\bauthor{\bsnm{Guo}, \binits{G.-C.}},
\bauthor{\bsnm{Gao}, \binits{X.}},
\bauthor{\bsnm{Pivoluska}, \binits{M.}},
\bauthor{\bsnm{Huber}, \binits{M.}}:
\batitle{Pathways for entanglement-based quantum communication in the face of high noise}.
\bjtitle{Phys. Rev. Lett.}
\bvolume{127},
\bfpage{110505}
(\byear{2021})
\doiurl{10.1103/PhysRevLett.127.110505}
\end{barticle}
\endbibitem

\bibitem[\protect\citeauthoryear{Hu et~al.}{2005}]{1377521}
\begin{barticle}
\bauthor{\bsnm{Hu}, \binits{X.-Y.}},
\bauthor{\bsnm{Eleftheriou}, \binits{E.}},
\bauthor{\bsnm{Arnold}, \binits{D.M.}}:
\batitle{Regular and irregular progressive edge-growth tanner graphs}.
\bjtitle{IEEE Transactions on Information Theory}
\bvolume{51}(\bissue{1}),
\bfpage{386}--\blpage{398}
(\byear{2005})
\doiurl{10.1109/TIT.2004.839541}
\end{barticle}
\endbibitem

\bibitem[\protect\citeauthoryear{Rusca et~al.}{2018}]{Rusca_2018}
\begin{barticle}
\bauthor{\bsnm{Rusca}, \binits{D.}},
\bauthor{\bsnm{Boaron}, \binits{A.}},
\bauthor{\bsnm{Gr\"unenfelder}, \binits{F.}},
\bauthor{\bsnm{Martin}, \binits{A.}},
\bauthor{\bsnm{Zbinden}, \binits{H.}}:
\batitle{Finite-key analysis for the 1-decoy state {QKD} protocol}.
\bjtitle{Applied Physics Letters}
\bvolume{112}(\bissue{17}),
\bfpage{171104}
(\byear{2018})
\doiurl{10.1063/1.5023340}
\end{barticle}
\endbibitem

\bibitem[\protect\citeauthoryear{Vagniluca et~al.}{2020}]{Vagniluca_2020}
\begin{botherref}
\oauthor{\bsnm{Vagniluca}, \binits{I.}},
\oauthor{\bsnm{Lio}, \binits{B.D.}},
\oauthor{\bsnm{Rusca}, \binits{D.}},
\oauthor{\bsnm{Cozzolino}, \binits{D.}},
\oauthor{\bsnm{Ding}, \binits{Y.}},
\oauthor{\bsnm{Zbinden}, \binits{H.}},
\oauthor{\bsnm{Zavatta}, \binits{A.}},
\oauthor{\bsnm{Oxenl{\o}we}, \binits{L.K.}},
\oauthor{\bsnm{Bacco}, \binits{D.}}:
Efficient time-bin encoding for practical high-dimensional quantum key distribution.
Physical Review Applied
\textbf{14}(1)
(2020)
\doiurl{10.1103/physrevapplied.14.014051}
\end{botherref}
\endbibitem

\bibitem[\protect\citeauthoryear{Zahidy et~al.}{2022}]{muji}
\begin{bchapter}
\bauthor{\bsnm{Zahidy}, \binits{M.}},
\bauthor{\bsnm{Ribezzo}, \binits{D.}},
\bauthor{\bsnm{De~Lazzari}, \binits{C.}},
\bauthor{\bsnm{Vagniluca}, \binits{I.}},
\bauthor{\bsnm{Biagi}, \binits{N.}},
\bauthor{\bsnm{Occhipinti}, \binits{T.}},
\bauthor{\bsnm{Oxenløwe}, \binits{L.K.}},
\bauthor{\bsnm{Galili}, \binits{M.}},
\bauthor{\bsnm{Hayashi}, \binits{T.}},
\bauthor{\bsnm{Antonelli}, \binits{C.}},
\bauthor{\bsnm{Mecozzi}, \binits{A.}},
\bauthor{\bsnm{Zavatta}, \binits{A.}},
\bauthor{\bsnm{Bacco}, \binits{D.}}:
\bctitle{4-dimensional quantum key distribution protocol over 52-km deployed multicore fibre}.
In: \bbtitle{2022 European Conference on Optical Communication (ECOC)},
pp. \bfpage{1}--\blpage{4}
(\byear{2022})
\end{bchapter}
\endbibitem

\bibitem[\protect\citeauthoryear{Declercq and Fossorier}{2007}]{4155118}
\begin{barticle}
\bauthor{\bsnm{Declercq}, \binits{D.}},
\bauthor{\bsnm{Fossorier}, \binits{M.}}:
\batitle{Decoding algorithms for nonbinary {LDPC} codes over {GF}$(q)$}.
\bjtitle{IEEE Transactions on Communications}
\bvolume{55}(\bissue{4}),
\bfpage{633}--\blpage{643}
(\byear{2007})
\doiurl{10.1109/TCOMM.2007.894088}
\end{barticle}
\endbibitem

\bibitem[\protect\citeauthoryear{Li et~al.}{2011}]{6125307}
\begin{bchapter}
\bauthor{\bsnm{Li}, \binits{E.}},
\bauthor{\bsnm{Gunnam}, \binits{K.}},
\bauthor{\bsnm{Declercq}, \binits{D.}}:
\bctitle{Trellis based extended {Min-Sum} for decoding nonbinary {LDPC} codes}.
In: \bbtitle{2011 8th International Symposium on Wireless Communication Systems},
pp. \bfpage{46}--\blpage{50}
(\byear{2011}).
\doiurl{10.1109/ISWCS.2011.6125307}
\end{bchapter}
\endbibitem

\bibitem[\protect\citeauthoryear{Xiao and Banihashemi}{2004}]{IPEG1}
\begin{barticle}
\bauthor{\bsnm{Xiao}, \binits{H.}},
\bauthor{\bsnm{Banihashemi}, \binits{A.H.}}:
\batitle{Improved progressive-edge-growth {(PEG)} construction of irregular {LDPC} codes}.
\bjtitle{IEEE Communications Letters}
\bvolume{8}(\bissue{12}),
\bfpage{715}--\blpage{717}
(\byear{2004})
\doiurl{10.1109/LCOMM.2004.839612}
\end{barticle}
\endbibitem

\bibitem[\protect\citeauthoryear{Venkiah et~al.}{2008}]{IPEG2}
\begin{bchapter}
\bauthor{\bsnm{Venkiah}, \binits{A.}},
\bauthor{\bsnm{Declercq}, \binits{D.}},
\bauthor{\bsnm{Poulliat}, \binits{C.}}:
\bctitle{Randomized progressive edge-growth {(RandPEG)}}.
In: \bbtitle{2008 5th International Symposium on Turbo Codes and Related Topics},
pp. \bfpage{283}--\blpage{287}
(\byear{2008}).
\doiurl{10.1109/TURBOCODING.2008.4658712}
\end{bchapter}
\endbibitem

\bibitem[\protect\citeauthoryear{Gorgoglione et~al.}{2010}]{5649264}
\begin{bchapter}
\bauthor{\bsnm{Gorgoglione}, \binits{M.}},
\bauthor{\bsnm{Savin}, \binits{V.}},
\bauthor{\bsnm{Declercq}, \binits{D.}}:
\bctitle{Optimized puncturing distributions for irregular non-binary {LDPC} codes}.
In: \bbtitle{2010 International Symposium On Information Theory and Its Applications},
pp. \bfpage{400}--\blpage{405}
(\byear{2010}).
\doiurl{10.1109/ISITA.2010.5649264}
\end{bchapter}
\endbibitem

\bibitem[\protect\citeauthoryear{Mitra et~al.}{2023}]{mitra2023nonbinary}
\begin{botherref}
\oauthor{\bsnm{Mitra}, \binits{D.}},
\oauthor{\bsnm{Tauz}, \binits{L.}},
\oauthor{\bsnm{Sarihan}, \binits{M.C.}},
\oauthor{\bsnm{Wong}, \binits{C.W.}},
\oauthor{\bsnm{Dolecek}, \binits{L.}}:
Non-Binary {LDPC} Code Design for Energy-Time Entanglement Quantum Key Distribution.
Arxive
(2023)
\end{botherref}
\endbibitem

\bibitem[\protect\citeauthoryear{Liu et~al.}{2022}]{Liu2022HighdimensionalQK}
\begin{botherref}
\oauthor{\bsnm{Liu}, \binits{J.-Y.}},
\oauthor{\bsnm{Lin}, \binits{Z.}},
\oauthor{\bsnm{Liu}, \binits{D.}},
\oauthor{\bsnm{Feng}, \binits{X.}},
\oauthor{\bsnm{Liu}, \binits{F.}},
\oauthor{\bsnm{Cui}, \binits{K.}},
\oauthor{\bsnm{Huang}, \binits{Y.}},
\oauthor{\bsnm{Zhang}, \binits{W.}}:
High-dimensional quantum key distribution using energy-time entanglement over 242 km partially deployed fiber.
Quantum Science and Technology
\textbf{9}
(2022)
\end{botherref}
\endbibitem

\bibitem[\protect\citeauthoryear{{Zhong} et~al.}{2015}]{2015NJPh...17b2002Z}
\begin{barticle}
\bauthor{\bsnm{{Zhong}}, \binits{T.}},
\bauthor{\bsnm{{Zhou}}, \binits{H.}},
\bauthor{\bsnm{{Horansky}}, \binits{R.D.}},
\bauthor{\bsnm{{Lee}}, \binits{C.}},
\bauthor{\bsnm{{Verma}}, \binits{V.B.}},
\bauthor{\bsnm{{Lita}}, \binits{A.E.}},
\bauthor{\bsnm{{Restelli}}, \binits{A.}},
\bauthor{\bsnm{{Bienfang}}, \binits{J.C.}},
\bauthor{\bsnm{{Mirin}}, \binits{R.P.}},
\bauthor{\bsnm{{Gerrits}}, \binits{T.}},
\bauthor{\bsnm{{Nam}}, \binits{S.W.}},
\bauthor{\bsnm{{Marsili}}, \binits{F.}},
\bauthor{\bsnm{{Shaw}}, \binits{M.D.}},
\bauthor{\bsnm{{Zhang}}, \binits{Z.}},
\bauthor{\bsnm{{Wang}}, \binits{L.}},
\bauthor{\bsnm{{Englund}}, \binits{D.}},
\bauthor{\bsnm{{Wornell}}, \binits{G.W.}},
\bauthor{\bsnm{{Shapiro}}, \binits{J.H.}},
\bauthor{\bsnm{{Wong}}, \binits{F.N.C.}}:
\batitle{{Photon-efficient quantum key distribution using time-energy entanglement with high-dimensional encoding}}.
\bjtitle{New Journal of Physics}
\bvolume{17}(\bissue{2}),
\bfpage{022002}
(\byear{2015})
\doiurl{10.1088/1367-2630/17/2/022002}
\end{barticle}
\endbibitem

\bibitem[\protect\citeauthoryear{Yang et~al.}{2021}]{yang2021fpga}
\begin{barticle}
\bauthor{\bsnm{Yang}, \binits{S.-S.}},
\bauthor{\bsnm{Liu}, \binits{J.-Q.}},
\bauthor{\bsnm{Lu}, \binits{Z.-G.}},
\bauthor{\bsnm{Bai}, \binits{Z.-L.}},
\bauthor{\bsnm{Wang}, \binits{X.-Y.}},
\bauthor{\bsnm{Li}, \binits{Y.-M.}}:
\batitle{An {FPGA}-based {LDPC} decoder with ultra-long codes for continuous-variable quantum key distribution}.
\bjtitle{IEEE Access}
\bvolume{9},
\bfpage{47687}--\blpage{47697}
(\byear{2021})
\end{barticle}
\endbibitem

\bibitem[\protect\citeauthoryear{Cui et~al.}{2013}]{6365815}
\begin{barticle}
\bauthor{\bsnm{Cui}, \binits{K.}},
\bauthor{\bsnm{Wang}, \binits{J.}},
\bauthor{\bsnm{Zhang}, \binits{H.-F.}},
\bauthor{\bsnm{Luo}, \binits{C.-L.}},
\bauthor{\bsnm{Jin}, \binits{G.}},
\bauthor{\bsnm{Chen}, \binits{T.-Y.}}:
\batitle{A real-time design based on {FPGA} for expeditious error reconciliation in {QKD} system}.
\bjtitle{IEEE Transactions on Information Forensics and Security}
\bvolume{8}(\bissue{1}),
\bfpage{184}--\blpage{190}
(\byear{2013})
\doiurl{10.1109/TIFS.2012.2228855}
\end{barticle}
\endbibitem

\end{thebibliography}
\appendix

\newpage

\section{Workflow of HD-Cascade}

\noindent HD-Cascade largely follows the description of binary Cascade in \cite{pacher} with the modifications described in Sec. \ref{hdc} and Sec. \ref{sec:phdc}. We give here a detailed description of the workflow of HD-Cascade for Bob's side, focusing on the more practical parallel implementation described in Sec. \ref{sec:phdc}. 

\noindent \textit{Preliminaries}: Two $q$-ary strings of length $n$, denoted by $\mathbf{x}$ and $\mathbf{y}$, and an estimate of the qudit error rate, denoted by $\text{QBER}_{\text{SYM}}$ in slight abuse of notation.
\vspace{2mm}

\subsection{General Flow} \label{sec:app:general_flow}

\noindent HD-Cascade, in general, follows an iteration pattern where for each iteration a permutation is applied. The key, i.e. the binary string, is then divided into blocks and a binary search locates errors. Each iteration is then followed by a Cascade step. \vspace{2mm}

\begin{enumerate}
    \item \textbf{Mapping:} Initially,  all symbols are mapped to an appropriate binary representation. Prior to the first iteration, all bits are shuffled while maintaining a record of which bits originate from the same symbol. These bits are denoted as ''partner bits". This mapping effectively reduces the expected QBER used for block-size calculations in all iterations to $\text{QBER}_{\text{BIN}} = \frac{q}{2(q-1)} \text{QBER}_{\text{SYM}}$. A storage container for keeping track of the smallest block and its parity value each bit takes part in for each iteration is initialized.\\

    \item \textbf{1st iteration:} In the first iteration,  the binary representation is split into $v=\log_2(q)$ groups of size $n$; $n$ being the number of qudits per frame. Each group only contains bits of the same bit-plane, i.e. the first group contains all the first bits of all symbols, the second group contains all the second bits of all symbols and so on. Permutations that are restricted to each group only are applied, i.e. we shuffle only inside each bit-plane. For the first group, the following is done:\\
    \begin{enumerate}
        \item  Divide the bits of the group into top-level blocks of size $k_1$,        
        \begin{equation} \label{eq:k_1}
            k_1 = \min (2^{[\log_2(1/\text{QBER}_{\text{BIN}})]},n/2)
        \end{equation}

        \noindent Calculate and compare the parities of these top-level blocks between Alice and Bob. Store the top-level blocks as the smallest block in iteration 1 for all bits inside blocks with matching parity. On blocks with mismatching parity, a binary search is executed to locate an error.\\
        
        \noindent \hspace{-0.6cm} \textbf{Binary Search:}
        \vspace{3mm}
        \begin{enumerate}
            \item Split the respective block in half.
            \item Calculate and exchange the parities of the two sub-blocks. One of the sub-blocks has a mismatching parity. Store the matching block as the smallest block for its bits in iteration 1.
            \item  If the mismatching sub-block contains only 1 bit, the error is found. Otherwise, repeat Step (a) with the mismatching sub-block.
        \end{enumerate}

    \vspace{4mm}
    \noindent After locating the error positions of one group using binary search, the corresponding partner bits can be located in all other groups. Request Alice to directly send the values of all partner bits and correct them. This reduces the expected QBER for all other groups. When calculating the top-level block sizes of the following blocks, adjust the expected QBER using the number of partner bits. It can be calculated for group $i$, $i=1,...,v$ as:

    \begin{equation}
        \text{QBER}_i = \text{QBER}_{\text{BIN}} - \frac{1}{2n}\sum_{j<i} \frac{\text{PB}_j}{v}, 
    \end{equation}

    \noindent where $\text{PB}_j$ denotes the number of partner bit requests originating from group $j$. Repeat step (a) for all other groups but replace $\text{QBER}_{\text{BIN}}$ with $\text{QBER}_i$. Collect all partner bits that are part of a group that has already been processed, e.g. some of the partner bits that occur when handling group $i=2$ but are part of group $i=1$. After completing all groups, use those partner bits as input for the Cascade step of the 1st iteration if their value differs between Bob and Alice. The Cascade step is described separately in Sec. \ref{sec:app:cas}\\
    \end{enumerate}
    
\item \textbf{2nd iteration:} The binary frame is again split into groups; this is also done in the binary protocol \cite{pacher}. The grouping follows the size $t$ of the smallest block each bit was part of in iteration 1, e.g. all bits whose top-level blocks matched are in one group, then all bits that were part of a block of half the top-level size that had matching parities between Alice and Bob, and so on. The QBER is adjusted according to the block size $t$ as \cite{pacher}:

\begin{align}
    \text{QBER}_{\text{2nd}}(t) &= \text{QBER}_{\text{BIN}} \frac{p_{\text{odd}}(t-1, \text{QBER}_{\text{BIN}})}{p_{\text{even}}(t,\text{QBER}_{\text{BIN}})}\\
    p_{\text{odd}}(t,p) &= \frac{1-(1-2p)^t}{2}\\
    p_{\text{even}}(t,p) &= \frac{1+(1-2p)^t}{2}
\end{align}

\noindent The top-level block size for each group is then calculated as 

\begin{equation} \label{eq:k_1}
    k_2 = \min (2^{[\log_2(2q/\text{QBER}_{\text{2nd}})]},n_j/2),
\end{equation}

\vspace{2mm}

\noindent with $n_j$ denoting the number of bits in group $j$. Proceed for each group as described for the first iteration but additionally keep track of all positions where an error is found by the binary search directly. Pass all relevant partner bits and error positions to the Cascade step.\\

\item \textbf{Remaining iterations:} The remaining iterations follow a more basic pattern. First, apply a random permutation to the binary frame. Divide the frame into top-level blocks of size $k_i$. As is common in binary Cascade, a doubling of the block sizes is done for each iteration, i.e. $k_{i+1} = 2k_{i}$, starting with $k_3=n/16$. The impact of the remaining iterations on the efficiency is rather low, as most errors can be expected to be corrected after the second iteration. Proceed for all blocks as described for iteration 1 and 2. Feed all known error positions to the Cascade step.
\end{enumerate}

\subsection{The Cascade step} \label{sec:app:cas}
The Cascade step expects a list of error positions as input. All those errors are already corrected, i.e. their bit value is flipped in Bob's current version of the key. 

\begin{enumerate}
    \item Go over all error positions and find the smallest block they take part in from all iterations that have not yet been processed. The term iteration refers to the iterations as described in Sec. \ref{sec:app:general_flow}. Note that each iteration applies a respective permutation. Pass the blocks to the next step.
    \newline Example: \textit{Consider the case where the bit at position 865 is incorrect. We check the smallest block information for this bit and see that it took part in a block of size 16, 8, and 4 in iterations 1,2, and 3, respectively. Iterations 2 and 3 are already flagged as processed for this bit, so proceed to the next step with the block of size 16 from iteration 1.}\vspace{2mm}
    \item For all blocks selected in Step 1, recalculate Bob's parity. Alice's parity is still known. If there is a mismatch, try to locate a new error by running a binary search on the blocks. This can be done either in parallel or in series. If done in parallel, multiple blocks originating from different iterations might point to the same error. Correct all found errors and request the respective partner bits from Alice. During the binary search new, smaller blocks are created. Update the smallest block information for all respective bits and iterations. Append all newly found errors, i.e. both from binary search and requested partner bits, to the list of known error positions. Mark the respective iterations as processed.
    \newline Example: \textit{We recalculate the parity of the block with size 16 from the previous step. Its parity now mismatches with that of Alice, and we perform a binary search on the block. As a result, a new error is found at position 35475. In the process, new blocks of size 8,4,2, and 1 with matching parity are created for iteration 1. The smallest block information is updated accordingly. We request the partner bits and receive the values of bits 35476 and 35477 of which only 35476 differs between Alice and Bob. Bit positions 35475 and 35476 are appended to the list of found errors. Iteration 1 is flagged as processed for bits 865 and 35475.} \vspace{2mm}

    \item Go over all error positions and check for each bit if all the smallest blocks it takes part in are already processed. If so, remove that position from the list. If the list of all error positions is empty, end the Cascade step, otherwise, repeat all steps. Example: \textit{All iterations are now flagged for bit 865 and it is therefore removed from the list of errors.}
    
\end{enumerate}

\end{document}